\documentclass[reprint,NumberedRefs]{JASA-MOD}

\usepackage{textcomp}
\newcommand{\mt}{\ensuremath{\times} }

\begin{document}

\title[Benchmark problems for transcranial ultrasound]{Benchmark problems for transcranial ultrasound simulation: Intercomparison of compressional wave models}

\author{Jean-Francois Aubry}
\thanks{Authors listed alphabetically}
\affiliation{Physics for Medicine Paris, INSERM U1273, ESPCI Paris, PSL University, CNRS UMR 8063}
\author{Oscar Bates}
\affiliation{Department of Bioengineering, Imperial College London, Exhibition Road, London, SW7 2AZ, United Kingdom}
\author{Christian Boehm}
\affiliation{Institute of Geophysics, ETH Z\"urich, Sonneggstrasse 5, 8092 Z\"urich, Switzerland}
\author{Kim Butts Pauly}
\affiliation{Department of Radiology, Stanford University, Stanford, CA, USA}
\author{Douglas Christensen}
\affiliation{Department of Biomedical Engineering and Department of Electrical and Computer Engineering, University of Utah}
\author{Carlos Cueto}
\affiliation{Department of Bioengineering, Imperial College London, Exhibition Road, London, SW7 2AZ, United Kingdom}
\author{Pierre G\'elat}
\affiliation{Department of Surgical Biotechnology, Division of Surgery and Interventional Science, University College London, London, NW3 2PF, United Kingdom}
\author{Lluis Guasch}
\affiliation{Earth Science and Engineering Department, Imperial College London, London, UK}
\author{Jiri Jaros}
\affiliation{Centre of Excellence IT4Innovations, Faculty of Information Technology, Brno University of Technology, Bozetechova 2, Brno, 612 00, Czech Republic}
\author{Yun Jing}
\affiliation{Graduate Program in Acoustics, The Pennsylvania State University, University Park, PA 16802, USA}
\author{Rebecca Jones}
\affiliation{Joint Department of Biomedical Engineering, University of North Carolina at Chapel Hill and North Carolina State University, NC, USA}
\author{Ningrui Li}
\affiliation{Department of Electrical Engineering, Stanford University, Stanford, CA, USA}
\author{Patrick Marty}
\affiliation{Institute of Geophysics, ETH Z\"urich, Sonneggstrasse 5, 8092 Z\"urich, Switzerland}
\author{Hazael Montanaro}
\affiliation{Foundation for Research on Information Technologies in Society (IT’IS), Zurich, Switzerland}
\affiliation{Laboratory for Acoustics / Noise control, Empa, Swiss Federal Laboratories for Materials Science and Technology, Dubendorf, Switzerland}
\author{Esra Neufeld}
\affiliation{Foundation for Research on Information Technologies in Society (IT’IS), Zurich, Switzerland}
\affiliation{Department of Information Technology and Electrical Engineering, Swiss Federal Institute of Technology (ETH), Zurich, Switzerland}
\author{Samuel Pichardo}
\affiliation{Radiology and Clinical Neurosciences Departments, Cumming School of Medicine, University of Calgary, Canada}
\author{Gianmarco Pinton}
\affiliation{Joint Department of Biomedical Engineering, University of North Carolina at Chapel Hill and North Carolina State University, NC, USA}
\author{Aki Pulkkinen}
\affiliation{Department of Applied Physics, University of Eastern Finland, 70211 Kuopio, Finland}
\author{Antonio Stanziola}
\affiliation{Department of Medical Physics and Biomedical Engineering, University College London, Gower Street, London, WC1E 6BT, United Kingdom}
\author{Axel Thielscher}
\affiliation{Technical University of Denmark}
\affiliation{Danish Research Center for Magnetic Resonance, Copenhagen University Hospital Hvidovre}
\author{Bradley Treeby}
\thanks{Corresponding author}
\email{b.treeby@ucl.ac.uk}
\affiliation{Department of Medical Physics and Biomedical Engineering, University College London, Gower Street, London, WC1E 6BT, United Kingdom}
\author{Elwin van 't Wout}
\affiliation{Institute for Mathematical and Computational Engineering, School of Engineering and Faculty of Mathematics, Pontificia Universidad Cat\'olica de Chile, Santiago, Chile}



\begin{abstract}
Computational models of acoustic wave propagation are frequently used in transcranial ultrasound therapy, for example, to calculate the intracranial pressure field or to calculate phase delays to correct for skull distortions. To allow intercomparison between the different modeling tools and techniques used by the community, an international working group was convened to formulate a set of numerical benchmarks. Here, these benchmarks are presented, along with intercomparison results. Nine different benchmarks of increasing geometric complexity are defined. These include a single-layer planar bone immersed in water, a multi-layer bone, and a whole skull. Two transducer configurations are considered (a focused bowl and a plane piston), giving a total of 18 permutations of the benchmarks. Eleven different modeling tools are used to compute the benchmark results. The models span a wide range of numerical techniques, including the finite-difference time-domain method, angular-spectrum method, pseudospectral method, boundary-element method, and spectral-element method. Good agreement is found between the models, particularly for the position, size, and magnitude of the acoustic focus within the skull. When comparing results for each model with every other model in a cross comparison, the median values for each benchmark for the difference in focal pressure and position are less than 10\% and 1 mm, respectively. The benchmark definitions, model results, and intercomparison codes are freely available to facilitate further comparisons.
\end{abstract}

\maketitle

\section{Introduction}

Ultrasound is increasingly used for therapeutic applications in the brain, including for tissue ablation,\cite{elias2016randomized} opening the blood-brain barrier,\cite{abrahao2019first} and for the modulation of brain activity.\cite{legon2014transcranial} One challenge is the non-invasive delivery of ultrasound through the skull bone, which can  significantly distort and attenuate the transmitted waves.\cite{hynynen1998demonstration} To account for this, computer simulations are now frequently used to make predictions of the intracranial pressure field,\cite{bouchoux2012experimental} and to correct for phase aberrations due to the skull.\cite{marquet2009non} This is particularly important for transcranial ultrasound stimulation (TUS), as the low ultrasound intensities make it highly challenging to measure the delivered energy in vivo, e.g., using MR-guided thermometry.\cite{dallapiazza2017noninvasive}

At a high level, there are four main steps in the setup of an acoustic model for transcranial ultrasound: (1) defining the medium parameters, including the skull and soft tissue geometry and the acoustic properties (using a medical image, for example); (2) defining the transducer characteristics, including the geometry, driving parameters, and relative position; (3) defining the numerical parameters for the model, including the grid resolution and boundary conditions; and (4) processing and interpreting the simulated results. One challenge for the community is that there is a large variation in these steps in the published literature, and there is currently little consensus on the best approach or the uncertainties associated with numerical modeling more generally.

As part of the International Transcranial Ultrasonic Stimulation Safety and Standards (ITRUSST) consortium, a working group focused on simulation and planning was convened. The primary goal was to perform a modeling intercomparison to systematically evaluate the steps involved in transcranial simulation, with a view to establishing best practice. A number of researchers active in the development of tools for transcranial ultrasound simulation were invited to take part. The first phase, which is reported here, was a model intercomparison using a series of numerical benchmarks relevant to transcranial ultrasound where the medium parameters and transducer characteristics were well defined. The primary research question was: \emph{do different modeling techniques and computer codes give the same answer when the inputs to the model are well specified?} This was taken as the first step to ensure that any differences in more complicated scenarios (e.g., where the skull properties are mapped from a medical image, or the transducer properties are mapped from a hydrophone measurement) could be evaluated as systematically and independently as possible.

The working group met regularly throughout 2021. The list of benchmarks (discussed in Sec.\ \ref{sec_benchmarks}) was iteratively refined, including the source definitions, the medium geometry, the material properties, and the output domain size. File submission formats, mechanisms for data sharing, and comparison metrics (along with codes to compute them\cite{code}) were also defined. Benchmark submissions were non-blinded with multiple resubmissions allowed. The goal was not a competition to establish which model was the “best” by some definition. Rather, it was to establish consensus on different approaches to transcranial ultrasound modeling, and how to implement these correctly using a range of modeling tools available to the community. In this spirit, work-in-progress  results and comparison metrics were discussed at regular intervals. These discussions, along with the sharing of code, approaches, and processing steps, etc, ultimately allowed the benchmarks to be computed with a wide set of simulation tools with excellent agreement (see Sec.\ \ref{sec_results}).

The primary goal of this phase of the intercomparison exercise was to establish a series of benchmarks relevant to transcranial ultrasound, along with consensus on the correct numerical solutions for these benchmarks. Consequently, simulations were typically performed with very high sampling to maximize accuracy. Because of this emphasis, and the different computational resources available to each group, a comparison of the computational performance of the individual models was considered out-of-scope from the outset. However, it is still important to note that some models used in the intercomparison, in particular those based on the angular spectrum method, have an efficiency/accuracy trade-off inherent in their formulation. This should be considered when interpreting the intercomparison metrics presented in Sec.\ \ref{sec_results} and the Supplementary Information.

The final output from the intercomparison exercise is a set of 9 well-defined numerical benchmarks relevant to transcranial ultrasound (with a total of 18 permutations of these benchmarks), along with publicly available simulation results for these benchmarks computed using 11 different modeling codes.\cite{results,code}

\section{\label{sec_benchmarks}Benchmarks}
\subsection{Overview}
The benchmarks were defined considering typical TUS scenarios, although they are also relevant to other applications of transcranial ultrasound. Simulations were single frequency (time-harmonic) and performed assuming linear wave propagation (the maximum pressure amplitude was less than 1.1 MPa for all benchmarks). Only compressional waves were considered, which is appropriate when the ultrasound waves are close to normal incidence to the skull bone.\cite{clement2004enhanced} All simulations were conducted in 3D.

\subsection{Transducer characteristics}
\begin{figure*}
\includegraphics[width=\textwidth]{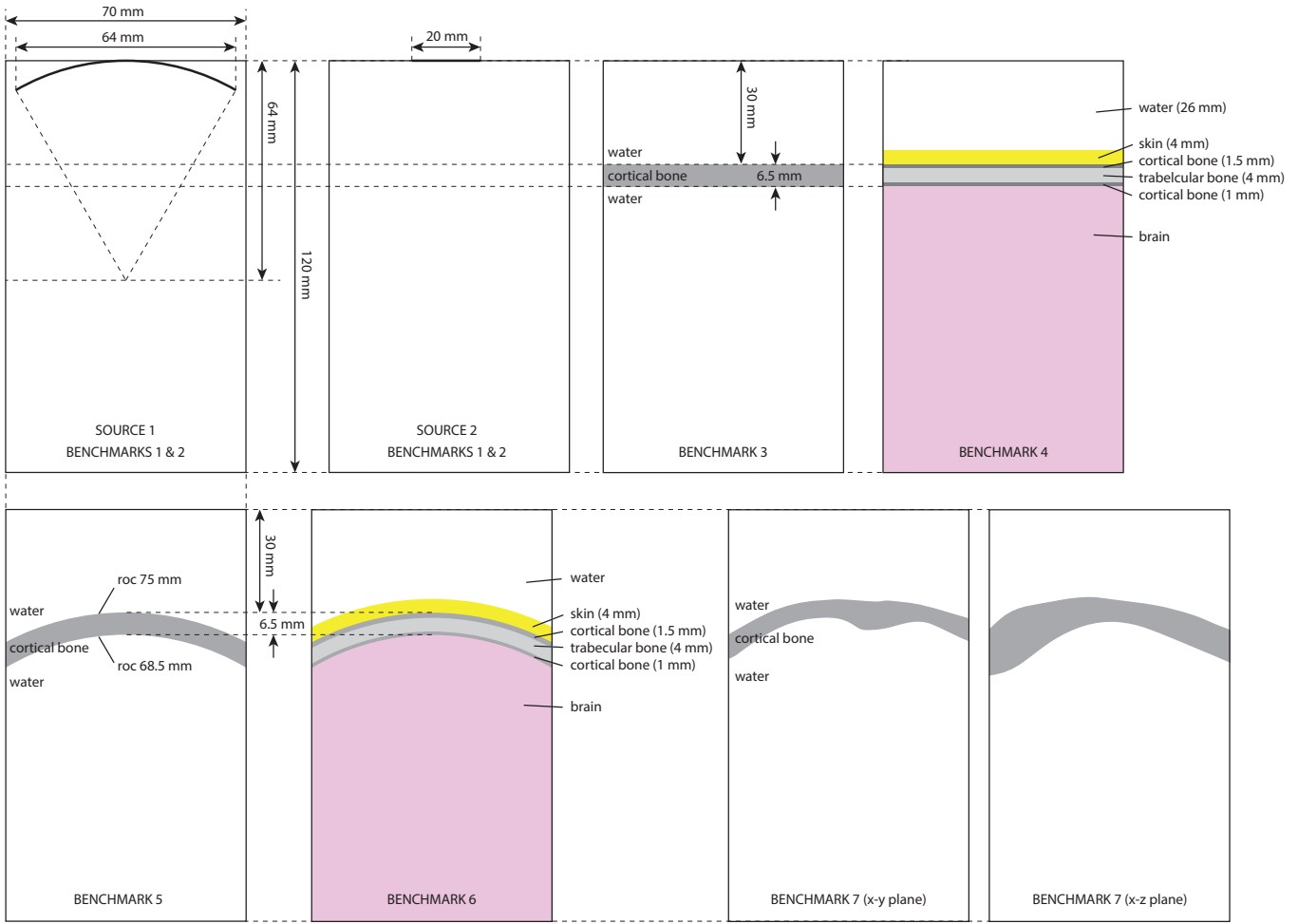}
\caption{\label{fig_benchmark_layouts}Transducer definitions and simulation layouts for benchmarks 1 to 7. Benchmarks 1 to 6 use a 2D comparison domain of 120 mm (axial) by 70 mm (lateral) through the central z-plane. Benchmark 7 uses a 3D comparison domain of 120 by 70 by 70 mm. The material properties used are given in Table \ref{tab_props}.}
\raggedright
\end{figure*}

Two transducer definitions were used (see Fig.\ \ref{fig_benchmark_layouts}). The first was a focused bowl transducer with a 64 mm radius of curvature and a 64 mm aperture diameter. This is representative of the single-element transducers frequently used for TUS.\cite{younan2013influence} The second was a plane piston transducer with a diameter of 20 mm. Piston transducers are often used in multi-element arrays. While the typical diameter of an element in a multi-element array is smaller than 20 mm, this diameter was used to provide identifiable beam characteristics within the simulation domain. For some numerical techniques, piston transducers are easier to model, particularly when aligned with the computational grid which avoids staircasing artifacts.\cite{robertson2017accurate} Both transducers were driven at 500 kHz with a constant surface velocity of 0.04 m/s.\cite{o1949theory} Assuming an acoustic impedance of 1.5 MRayls, this is equivalent to modeling the sources as a distribution of free-field monopole radiators with a source pressure of 60 kPa.

\subsection{Material properties}
The material properties used for the benchmarks are given in Table \ref{tab_props}. These are intended to be representative (rather than definitive) values, and were taken from the range presented in the literature.\cite{fry1978acoustical,mast2000empirical,clement2002correlation,pichardo2010multi,pinton2012attenuation,pichardo2017viscoelastic,white2006longitudinal,webb2020acoustic} For the simulations including absorption, the loss is defined to be non-dispersive, i.e., either frequency independent, or for power law models, dependent on frequency squared.

\begin{table}[!t]
\caption{\label{tab_props}Compressional sound speed ($c$), mass density ($\rho$), and absorption coefficient ($\alpha$) used in the benchmark simulations.}

\begin{ruledtabular}
\begin{tabular}{lccc}
 & $c$ (m/s) & $\rho$ (kg/m$^3$) & $\alpha$ (dB/cm at 500 kHz) \\
\hline
Water & 1500 & 1000 & 0 \\
Skin & 1610 & 1090 & 0.2 \\
Brain & 1560 & 1040 & 0.3 \\
Cortical Bone & 2800 & 1850 & 4 \\
Trabecular Bone & 2300 & 1700 & 8 \\
\end{tabular}
\end{ruledtabular}
\end{table}

\subsection{\label{sec_simulation_outputs}Simulation outputs}
The simulation results were stored as two variables named \verb=p_amp= and \verb=p_phase=. These represent the amplitude and phase of the complex pressure field at 500 kHz over the specified comparison domain. For time domain models, these parameters can be extracted precisely by setting the time step to an integer number of points per period (PPP), recording the steady state pressure field for an integer number of periods, and then extracting the amplitude and phase at the driving frequency using a Fourier transform. Note, the phase is optional and was not used for the comparisons presented in Sec.\ \ref{sec_results}, but was included for completeness. The results were saved either as MATLAB \verb=.mat= files using the \verb=`-v7.3'= flag where possible (this format can be easily opened as a HDF5 file outside MATLAB), or as HDF5 \verb=.h5= files with the variables saved as datasets in the root group.

Regardless of the sampling or mesh used for the simulations, the outputs stored in the comparison files were re-sampled onto a uniform Cartesian grid with 0.5 mm grid sampling. This corresponds to six points per wavelength (PPW) in water. For benchmarks 1 to 6, the comparison domain size was a 120 \mt 70 mm (axial \mt lateral) slice through the central z-coordinate, corresponding to a grid size of 241 \mt 141 grid points. For benchmark 7, the comparison domain was 120 \mt 70 \mt 70 mm (241 \mt 141 \mt 141 grid points). For benchmark 8, the comparison domain was 225 \mt 170 \mt 190 mm (451 \mt 341 \mt 381 grid points). For benchmark 9, the comparison domain was 212 \mt 224 \mt 184 mm (425 \mt 449 \mt 369 grid points).

For all benchmarks, the transducer was oriented such that the beam axis pointed in the x-dimension, with the transducer positioned in the center of the y/z-dimensions. Using 1-based indexing, the center of the source (rear of the bowl or center of the piston) relative to the output grid was positioned at \verb=[1, 71]= for benchmarks 1 to 6, \verb=[1, 71, 71]= for benchmark 7, \verb=[1, 171, 191]= for benchmark 8, and \verb=[1, 225, 185]= for benchmark 9. Note, the comparisons for benchmarks 1 to 6 were made in 2D due to the axisymmetry of the geometry. All simulations were conducted in 3D.

\subsection{Naming convention}
The benchmarks were given unique identifiers in the following format: \verb=PH<NUM>-BM<NUM>-SC<NUM>=. \verb=PH= (phase) identifies the intercomparison phase (in this case 1). \verb=BM= (benchmark) identifies the benchmark number within the phase. \verb=SC= (source) identifies the source condition, where 1 is the bowl source, and 2 is the plane piston source. A summary of the different benchmarks is given in Table \ref{tab_benchmarks}. File names for the intercomparison results follow the same convention with the model name appended (see Table \ref{tab_models}): \verb=PH<NUM>-BM<NUM>-SC<NUM>_<MODELNAME>=. The simulation outputs for each model for each benchmark are publicly available.\cite{results}

\begin{table*}[tbh]
\caption{\label{tab_benchmarks}Summary of benchmarks in Phase 1 of the intercomparison. SC1 corresponds to the focused bowl transducer and SC2 the plane piston transducer. Outputs are resampled to a regular Cartesian mesh with a grid spacing of 0.5 mm. Simulation layouts are shown in Figs.\ \ref{fig_benchmark_layouts} and \ref{fig_benchmark_layouts_78}. gp = grid points.}

\begin{ruledtabular}
\begin{tabular}{llll}
Label & Description & Output Grid Size \\
\hline
\verb=PH1-BM1-SC1/2= & Water (lossless) & 120$\times$70 mm (241$\times$141 gp)\\
\verb=PH1-BM2-SC1/2= & Water (artificial absorption of 1 dB/cm at 500 kHz) & 120$\times$70 mm (241$\times$141 gp)\\
\verb=PH1-BM3-SC1/2= & Flat, single-layer skull (cortical bone) in water & 120$\times$70 mm (241$\times$141 gp)\\
\verb=PH1-BM4-SC1/2= & Flat, skin, three-layered skull, and brain & 120$\times$70 mm (241$\times$141 gp)\\
\verb=PH1-BM5-SC1/2= & Curved, single-layer skull (cortical bone) in water & 120$\times$70 mm (241$\times$141 gp)\\
\verb=PH1-BM6-SC1/2= & Curved, skin, three-layered skull, and brain & 120$\times$70 mm (241$\times$141 gp)\\
\verb=PH1-BM7-SC1/2= & Truncated skull mesh in water, target in visual cortex & 120$\times$70$\times$70 mm (241$\times$141$\times$141 gp)\\
\verb=PH1-BM8-SC1/2= & Whole skull mesh, target in visual cortex & 225$\times$170$\times$190 mm (451$\times$341$\times$381 gp)\\
\verb=PH1-BM9-SC1/2= & Whole skull mesh, target in motor cortex & 212$\times$224$\times$184 mm (425$\times$449$\times$369 gp)\\
\end{tabular}
\end{ruledtabular}
\end{table*}

\subsection{\label{sec_benchmark_list}Benchmarks}
\begin{figure}
\includegraphics[width=\reprintcolumnwidth]{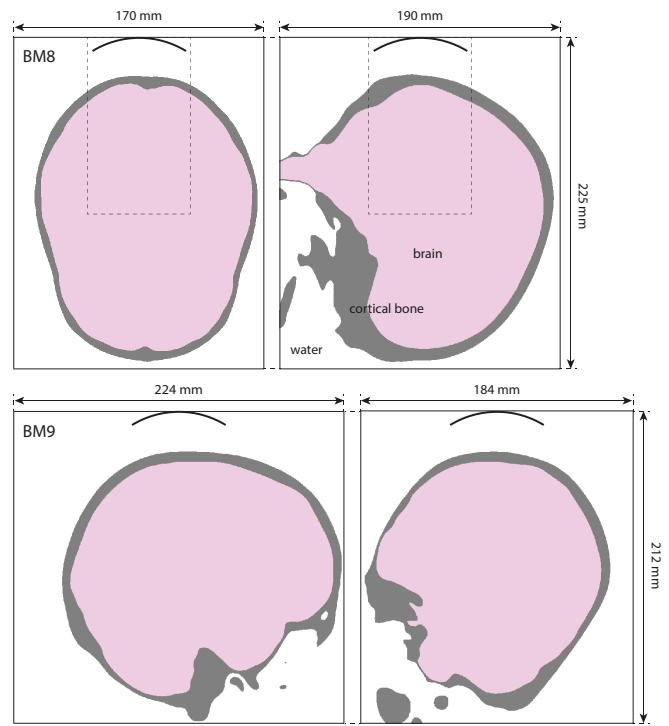}
\caption{\label{fig_benchmark_layouts_78}Simulation layouts for benchmarks 8 (top row) and 9 (bottom row) showing the central x-y and x-z slices. The position of the bowl transducer is shown for reference. Benchmark 7 (shown in Fig.\ \ref{fig_benchmark_layouts}) uses a subset of the skull mask and the same relative transducer position as benchmark 8, with a reduced comparison domain size as shown with the dashed line. The material properties used are given in Table \ref{tab_props}.}
\raggedright
\end{figure}

A total of 9 benchmarks relevant to transcranial ultrasound were devised. These are summarized in Table \ref{tab_benchmarks}. The benchmarks gradually increase in complexity, adding both additional tissue layers, and increasing the geometric complexity of the skull. Benchmarks 1 to 7 are illustrated in Fig.\ \ref{fig_benchmark_layouts} while benchmarks 8 and 9 are illustrated in Fig.\ \ref{fig_benchmark_layouts_78}.

Benchmark 1 considers the bowl and piston transducers in water (free-field) using the properties given in Table \ref{tab_props}. Benchmark 2 adds uniform artificial absorption of 1 dB/cm at 500 kHz. During the intercomparison exercise, these benchmarks served as a helpful reference to ensure the transducer properties, absorption units, and comparison domain were correctly specified. For these simulations, reference simulations were also computed using the fast near-field method as implemented in the FOCUS toolbox.\cite{mcgough2004efficient,chen20082d,kelly2009transient} Calculations using FOCUS were performed using 5000 integration points to give a high level of precision. Several models used the fields computed using FOCUS across a transverse y-z plane as the source definition (see Sec.\ \ref{sec_models}).

Benchmark 3 introduces a single flat 6.5 mm layer of cortical bone immersed in water, positioned 30 mm from the transducer as shown in Fig.\ \ref{fig_benchmark_layouts}. Benchmark 4 extends this to include a 4 mm skin layer, a three-layered skull (consisting of 1.5 mm cortical bone for the outer table, 4 mm trabecular bone, and 1 mm cortical bone for the inner table, giving the same overall skull thickness and position as benchmark 3), with water on the exterior and brain on the interior as shown in Fig.\ \ref{fig_benchmark_layouts}. The thickness values are based on average values for parietal bone\cite{alexander2019structural} and scalp.\cite{hori1972thickness}

Benchmark 5 increases the geometric complexity of benchmark 3 by using a curved 6.5 mm layer of cortical bone immersed in water, with inner and outer radii of 68.5 mm and 75 mm, respectively. Note, the bone layer is spherically (not cylindrically) curved, meaning the curvature in the out-of-plane dimension is the same as that shown in Fig.\ \ref{fig_benchmark_layouts}. Benchmark 6 is a curved extension of benchmark 4, where the thickness values correspond to differences in the curvature radii.

Benchmarks 7 to 9 increase the geometric complexity further by using a homogeneous skull mesh generated from the \verb=MNI152_T1_1mm= magnetic resonance imaging template brain.\cite{fonov2009unbiased,fonov2011unbiased} The template image was run through an adapted version of SimNIBS \verb=headreco=.\cite{nielsen2018automatic} Additional smoothing of the tissue surfaces while simultaneously preventing intersections between neighboring surfaces was performed using SimNIBS functions. Benchmarks 7 and 8 use a transducer position targeted at the foveal representation of the primary visual cortex, while benchmark 9 uses a transducer position targeted at the hand area of the primary motor cortex. 

The skull mesh was stored as two \verb=.stl= files representing the inner and outer surfaces of the skull bone. Position transforms were stored as 3D affine transformations which position the transducer relative to the coordinates in the \verb=.stl= files. Grid-based discretizations containing a binary skull mask were also generated using the \verb=iso2mesh= MATLAB toolbox.\cite{iso2mesh,fang2009tetrahedral} These were generated on a regular Cartesian mesh at a range of resolutions after applying the appropriate inverse position transforms (to move the skull mesh relative to the transducer), and were truncated to the appropriate comparison domain (see Sec.\ \ref{sec_simulation_outputs}). The skull files and position transforms are available alongside the simulations results.\cite{results}

\subsection{\label{sec_metrics}Intercomparison metrics}

A number of metrics were chosen to compare the simulated fields. Mathematical definitions for some metrics are given in Table \ref{tab_error_metrics}. Metrics based on the entire field were taken from the exit plane of the source, excluding the first grid point in the x-direction for the piston transducer, and the first 19 grid points in the x-direction for the bowl transducer. The relative $L^2$ and $L^\infty$ errors provide a useful (and strict) measure of the overall differences between simulations. However, for more complex geometries, these become dominated by differences in the rapidly-varying near-field region between the source and the skull. For this reason, differences in the focal characteristics were also computed. This included the magnitude and position of the peak pressure within the brain, and differences in the full-width at half-maximum (FWHM) and -6dB focal volume. The FWHM values were taken in each Cartesian direction present in the comparison domain (i.e., in the $x$ and $y$ dimensions for benchmarks 1 to 6, and in the $x$, $y$, and $z$ dimensions for benchmarks 7 to 9). For benchmarks 3 to 9 for the piston transducer, the acoustic field gradually decays within the brain, thus there is no natural focus in the axial direction. In this case, the axial focal position, and lateral profiles and FWHM values were taken at $x = 60$ mm (corresponding to a grid index of 121). For benchmarks 7 to 9, differences in the -6 dB focal volume were also computed. The focal volume was calculated by thresholding the pressure field inside the brain to 50\% of the maximum value, and then counting the voxels in the largest connected component. Code to compute the intercomparison metrics is available on GitHub.\cite{code}

\begin{table}[!t]
\caption{\label{tab_error_metrics}Difference metrics used for the intercomparison. Here $p_1$ and $p_2$ are the amplitude of the pressure field over the 2D or 3D comparison domains for the reference field and comparison field, respectively (these are assumed to be positive). Sums and maximum values are assumed to be over all values in the comparison domain starting from the exit plane of the transducer. Focal values are taken from inside the brain (or post-skull) region only. $\text{pos max}$ is used to denote the position of the maximum value in the comparison domain.}

\begin{ruledtabular}
\renewcommand{\arraystretch}{1.5}
\begin{tabular}{lp{5cm}}
Metric & Definition \\
\hline
Relative $L^2$ & $\sqrt{\frac{\sum{(p_1 - p_2)^2}}{\sum{p_1^2}}}$ \\
Relative $L^\infty$  & $\frac{\max{\left| p_1 - p_2 \right|}}{\max{\left( p_1\right)}}$\\
Focal (peak) pressure & $\frac{\left|\max{\left( p_1 \right)} - \max{\left(p_2 \right)}\right|}{\max{\left( p_1\right)}}$\\
Focal position & $\| \text{pos max}(p_1) - \text{pos max}(p_2) \|_2$ \\
\end{tabular}
\end{ruledtabular}
\end{table}

\section{\label{sec_models}Models}

\subsection{Overview}

A total of 11 modeling tools were used for the intercomparison, in addition to the free-field reference values calculated using FOCUS discussed in Sec.\ \ref{sec_benchmark_list}. These are summarized in Table \ref{tab_models}. A short description of each model is given in the following sections, with additional details given in the Supplementary Material.

\begin{table*}[!t]
\caption{\label{tab_models}Summary of models used to calculate the benchmark results. Additional details given in the Supplementary Material. Authors correspond to the authors of the current manuscript directly contributing to the intercomparison exercise, not necessarily the authors of the model.}

\begin{ruledtabular}
\begin{tabular}{lccc}
Label & Authors & Domain & Method
 \\
\hline
BABELVISCOFDTD & SP & Time Domain & Finite-Difference Time-Domain \\
FULLWAVE & RJ, GP &  Time Domain & Finite-Difference Time-Domain \\
GMFDTD & AP & Time Domain & Finite-Difference Time-Domain \\
HAS & NL, KBP & Frequency Domain & Hybrid Angular Spectrum \\
JWAVE & AS & Frequency Domain & Fourier Spectral Method With Iterative Solver\\
KWAVE & BT, JJ & Time Domain & Pseudo-spectral Time Domain \\
MSOUND & YJ & Frequency Domain & Modified Angular Spectrum\\
OPTIMUS & PG, EvW & Frequency Domain & Boundary Element Method\\
SALVUS & PM, CB & Time Domain & Spectral Element Method\\
SIM4LIFE & HM, EN & Time Domain & Finite-Difference Time-Domain\\
STRIDE & CC, OB, LG & Time Domain & Finite-Difference Time-Domain\\
\end{tabular}
\end{ruledtabular}
\end{table*}

\subsection{BABELVISCOFDTD}

BabelViscoFDTD solves the viscoelastic wave equation expressed in stress tensors and displacement vectors, where the bone material is modeled as a viscoelastic isotropic medium.\cite{pichardo2017viscoelastic} The term “Babel” refers to the multiple computing backends (CUDA, OpenCL, Apple Metal and X86-64) that are supported for calculations. Nodes of stress and displacement are placed in a staggered grid arrangement.\cite{virieux1986p} Calculations are solved using a 4th-order in space and 2nd-order in time finite-difference time-domain (FDTD) scheme in Cartesian coordinates.\cite{levander1988fourth,bohlen2002parallel}  Stress tensors and displacement vectors are solved a half time step separated from each other. Attenuation losses are modeled using a quality factor for narrowband conditions.\cite{blanch1995modeling,bohlen2002parallel} Liquid-bone interfaces and heterogeneity of tissue material are modeled using averaging operators.\cite{moczo20023d} Optional reduction of staircasing artifacts can be enabled using a superposition operator.\cite{drainville2019superposition} A perfectly matched layer (PML) condition for viscoelastic propagation is used to absorb waves at the boundaries.\cite{collino2001application}

All benchmarks were computed using a resolution of 12 grid PPW. Sources were modeled as stress nodes using the same staircase-free formulation and dispersion correction as in the k-Wave model (see Sec.\ \ref{sec_kwave}).  The time PPP for benchmarks 1 and 2 was 25, and 48 for benchmarks 3 to 9. Benchmarks 1 to 7 used a total grid size of 305 $\times$ 305 $\times$ 521 grid points including the PML. For benchmarks 8 and 9 the grid size was, respectively, 785 $\times$ 705 $\times$ 941 and 761 $\times$ 921 $\times$ 889. Simulation outputs were resampled to the comparison grid using a spline interpolation of order 3.

\subsection{FULLWAVE}

Fullwave2 3D solves the wave equation with quadratic nonlinearity and multiple relaxations using a staggered-grid FDTD approach with fourth-order accuracy in time and variable accuracy in space.\cite{pinton2009heterogeneous,pinton2021fullwave} This model uses a staggered-grid Cartesian mesh with a convolutional PML at the boundaries, utilizing high-order adaptive stencils that minimize dispersion and dissipation errors. The source and output can take the shape of any arbitrary geometry that can be defined on a Cartesian grid, with sources modeled either as free-field particle displacement, velocity, or a monopole pressure source.

For the benchmark comparison, the bowl and piston geometries were modeled as monopole pressure sources on a Cartesian grid, emitting a continuous sinusoidal wave. All benchmarks were computed with 12 PPW and 60 PPP, giving a Courant-Fredrichs-Lewy condition (CFL)\cite{courant1928partiellen} of 0.2. This created a simulation grid 2$\times$ the size of the comparison grid. To account for this, the simulations were run with a spatial step size of 2 voxels in each direction, downsampling the output grid to the comparison grid size. The output over one steady-state cycle was then scaled based on the CFL and driving signal to account for the use of additive sources.

\subsection{GMFDTD}

The GMFDTD model simulates acoustic wave-propagation based on coupling of the second-order acoustic and viscoelastic wave-equations using a combined grid method and FDTD method. The model operates using a regular Cartesian mesh. For fluid simulations (as described in this work), GMFDTD solves the acoustic wave-equation using a FDTD approach with fourth-order spatial and second-order time stencils. First-order absorbing boundary conditions are used on exterior boundaries of the simulation domain. A finite thickness absorbing layer was placed on the exterior boundaries to further reduce acoustic reflections. A heterogeneous Neumann boundary condition is used to model the sound sources making the source-medium interface work as an acoustically hard reflector for incoming sound waves. For a more thorough description of the model, see Ref. \cite{pulkkinen2014numerical}.

All simulations were computed using 12 PPW and 75 PPP. Grid sizes for the simulations were 576 $\times$ 376 $\times$ 376 grid points for benchmarks 1 to 7, 996 $\times$ 776 $\times$ 856 for benchmark 8, and 944 $\times$ 992 $\times$ 832 for benchmark 9. The grid sizes include a 48 grid point absorbing layer surrounding the domain, which had attenuation linearly increasing from zero to 50 Np/m corresponding to about 94\% amplitude attenuation for a normally incident reflected wave. Simulations were computed for 3,900 time steps for benchmarks 1 to 7, 15,075 time steps for benchmark 8, and 15,075 time steps for benchmark 9. The simulations produced a complex valued steady state pressure field which was resampled to the comparison grid using spline interpolation before computing the pressure amplitude and the phase angle.

\subsection{HAS}

The hybrid angular spectrum (HAS) method is a generalization of the angular spectrum method, enabling propagation of pressure fields in heterogeneous media.\cite{vyas2012ultrasound,almquist2016rapid} An initial pressure distribution is first defined on a plane perpendicular to the direction of propagation. To produce the full 3D steady-state pressure field, pressures on subsequent planes are calculated in the spatial-frequency domain by solving the Helmholtz equation using the angular spectrum method. Errors due to local variations in attenuation and acoustic velocity are corrected for using a spatial step between each spatial-frequency step. Reflected pressures are saved, backpropagated, and summed with the incident pressure field, and this process is repeated until convergence to produce the final steady-state pressure field.

Initial pressure fields were computed using the fast near-field method as implemented in the FOCUS toolbox (see Sec.\ \ref{sec_benchmark_list}). Benchmarks 1 to 6 were computed using a grid size of $1001 \times 1001 \times 1001$ with 6 PPW in the transverse directions and 24 PPW in the axial direction. Benchmarks 7 and 8 were computed using a grid size of $1401 \times 1401 \times 501$ with an isotropic resolution of 12 PPW. Benchmark 9 was computed using a grid size of $1201 \times 1201 \times 501$ with an isotropic resolution of 12 PPW. Calculated pressure fields were resampled to the comparison grid using bilateral interpolation.

\subsection{JWAVE}

JWAVE simulates the solution of time-harmonic wave propagation problems by solving the heterogeneous Helmholtz equation in the complex domain, using a regular Fourier spectral collocation method and linear iterative solvers such as restarted GMRES.\cite{saad1986gmres} Absorbing boundary conditions are enforced using a PML,\cite{bermudez2007optimal} while the definition of the sources is done by projecting them on the discrete collocation grid by approximately convolving them with the band-limited interpolant.\cite{wise2019representing} The source field is modeled as a mass source. JWAVE is a Python software written using JaxDF,\cite{stanziola2021research} which in turn is based on JAX.\cite{jax2018github} The code is just-in-time compiled for the hardware at hand (e.g. GPUs or TPUs) and allows for automatic differentiation to be applied with respect to any continuous parameter.

Benchmarks 1 and 2 were computed using 6 PPW, while benchmarks 3 to 7 were computed using 12 PPW. The PML size was fixed to 30 voxels. To reduce the computation time of the FFTs, the domain dimensions were padded to the nearest integers with prime factors smaller than 7. When required, the results were resampled to the intercomparison grid using Fourier interpolation. Benchmarks 8 and 9 were too large for the available computational resources, so results for these benchmarks were not computed.

\subsection{\label{sec_kwave}KWAVE}

k-Wave solves three coupled equations equivalent to a generalized Westervelt equation, where spatial gradients are calculated using a Fourier collocation spectral method, and time integration is performed using a dispersion-corrected finite-difference scheme.\cite{treeby2010k,treeby2012modeling} Calculations are performed on a regular Cartesian mesh with a space and time staggered grid. A split-field PML is used to absorb the waves at the domain boundaries. Sources are modeled as free-field monopoles (injection of mass) using a staircase-free formulation to represent the bowl and piston geometries,\cite{wise2019representing} and a dispersion-corrected time-stepping scheme.\cite{cox2018accurate}

Benchmarks 1 to 6 were computed using the axisymmetric version of k-Wave to provide a high-resolution reference simulation.\cite{treeby2020nonlinear} Benchmarks 1, 2, 3, and 5 used 60 PPW and 2400 PPP, while benchmarks 4 and 6 used 60 PPW and 6000 PPP. In both cases, the total grid size was 2700 $\times$ 864 grid points including the PML and the simulation time was 120 $\mu$s, giving 144,000 and 360,000 time steps, respectively. Benchmarks 7 to 9 were computed using the 3D version of k-Wave optimized for high-performance computing clusters.\cite{jaros2016full} Benchmark 7 used 30 PPW and 1200 PPP, with a grid size of 1296 $\times$ 768 $\times$ 768 grid points and 72,000 time steps (120 $\mu$s simulation time). Benchmarks 8 and 9 used 18 PPW and 360 PPP with 72,000 time steps (400 $\mu$s simulation time). The grid sizes were 1458 $\times$ 1080 $\times$ 1200 and 1350 $\times$ 1440 $\times$ 1152, respectively. The simulation times were sufficient to reach steady state and were chosen via a convergence test. All simulations used a grid spacing that was an integer division of the comparison resolution (0.5 mm), thus simulations outputs were resampled to the comparison grid using decimation.

\subsection{MSOUND}

mSOUND solves the Helmholtz equation with the absorption term for linear acoustics cases.\cite{gu2021msound} For layered media, the conventional angular spectrum approach coupled with the analytical plane wave transmission and reflection coefficients is used. For arbitrarily heterogeneous media, a split-step Fourier method with interpolation is used. Calculations are performed on a regular Cartesian mesh in space. A non-reflecting layer can be used to reduce the spatial aliasing error. Sources are modeled by assigning the complex pressure distribution on the initial plane. In these simulations, the initial plane pressure fields were obtained by FOCUS, as mSOUND currently only considers the pressure-release boundary condition (p=0) for the region outside the source. All benchmarks were computed using the function \verb=Forward3D_fund=. Benchmark 1 and 2 were computed using 6 PPW in all directions. Benchmarks 7 to 9 were computed using 12 PPW in all directions. Benchmarks 3 and 4 were computed using 6 PPW in the lateral directions, and 48 PPW in the axial (propagation) direction. Benchmarks 5 and 6 were computed using 6 PPW in the lateral directions, and 24 PPW in the axial direction. For benchmarks 3 to 9, simulation outputs were down-sampled to the comparison grid.

\subsection{OPTIMUS}

OptimUS is a full wave solver based on the boundary element method (BEM).\cite{wout2015fast} The BEM employs the Green's function of the Helmholtz equation to reformulate the volumetric wave problem into a boundary integral equation at the interfaces of piecewise homogeneous domains embedded in free space.\cite{wout2022highcontrast} Benchmarks 3, 5 and 7 were modeled with the PMCHWT formulation,\cite{haqshenas2021fast} benchmarks 4 and 6 were solved with a multi-trace formulation,\cite{wout2021benchmarking} and a nested version of the PMCHWT formulation solves benchmarks 8 and 9. The numerical discretization leads to a dense system of linear equations, whose computational footprint is reduced through hierarchical matrix compression.\cite{betcke2017computationally} The convergence of the iterative GMRES linear solver was improved with OSRC preconditioning.\cite{wout2021pmchwt} All models were implemented in Python, using version 3 of the open-source BEMPP library.\cite{smigaj2015solving} The triangular surface meshes were created with Gmsh~\cite{geuzaine2009gmsh} for benchmarks 3 to 6 and using Meshmixer~\cite{meshmixer} for benchmarks 7 to 9.

The size of the mesh elements was specified as 4.3 PPW (0.7~mm) in benchmark 3, 6 PPW (0.5~mm) in benchmarks 4 and 5, 4 PPW (0.75 mm) for benchmark 6, and 10 PPW (0.3~mm) for benchmark 7. A compromise in terms of memory requirements and accuracy of results had to be sought on benchmarks 8 and 9, and a value of 4 PPW (0.75~mm) was used on the skull mesh in the vicinity of the transducer with a value of 2.4 PPW (1.25~mm) elsewhere. The bowl and piston transducers were implemented using a Rayleigh integral formulation, consisting of a summation of evenly spaced monopole radiators positioned on their surface. The transducer surfaces were discretized using 23 and 6 monopole sources per wavelength in water for benchmarks 1 to 7 and benchmarks 8 and 9, respectively. In cases where the position of monopole sources coincided with a field evaluation point, NaN was assigned to the acoustic pressure. The acoustic field was evaluated from the surface potentials by interpolation for points on, or very close to, the material interface and with Green's functions for points in the material volume.

\subsection{SALVUS}

Salvus solves the second-order linear wave equation in the time-domain and can handle acoustic and elastic media.\cite{afanasiev2019modular} It utilizes a matrix-free implementation of the continuous-Galerkin spectral-element method\cite{ferroni2017dispersion} and an explicit second-order Newmark time-stepping scheme. The computational domain is discretized using unstructured conforming hexahedral meshes,\cite{hapla2021fully} which enable the exact representation of interfaces and discontinuities in the tissue parameters.  Absorbing boundaries are imposed using the first-order Sommerfeld radiation condition in addition to sponge layers.\cite{kosloff1986absorbing} The transducers are modeled as a collection of monopole point sources distributed evenly over the surface of the transducer.
 
Spectral elements of order 4 were utilized for all simulations; this corresponds to 125 nodes per element. Due to the interfaces being represented precisely using hexahedral meshes generated within Coreform Cubit 2021.5,\footnote{Coreform Cubit (Version 2021.5) [Computer software]. Orem, UT: Coreform LLC. Retrieved from http://coreform.com} utilizing 2 to 3 elements per wavelength for all benchmarks proved to be sufficient. The maximum pressure distributions were computed by propagating the wavefield in the time domain and then applying the on-the-fly temporal Fourier transform.\cite{witte2019compressive}  All simulation results were output on the same hexahedral discretizations used as inputs and were subsequently resampled onto the comparison grid using fourth-order Lagrange polynomials in the spectral-element basis.

\subsection{SIM4LIFE}

Sim4Life solves acoustic pressure wave equations (linear, or Westervelt-Lighthill, which considers dispersion and frequency mixing), using a multi-GPU-accelerated FDTD method on adaptive, rectilinear meshes (to adapt grid-steps to the local wavelength and refine relevant geometric features) with cell-centered pressure degrees-of-freedom. Flux conserving virtual auxiliary points are used to improve accuracy at interfaces and boundaries, and PMLs -- according to the stretched coordinate formulation \cite{chew19943d} -- are used to avoid reflections at domain boundaries (for more details on the numerical methods, see  \cite{kyriakou2015full}). Results can be recorded as phasors (at the base frequency and, if relevant, higher harmonics) or transient 3 + 1D fields, and the solver has been verified and validated,\cite{ neufeld2016approach} also for transcranial focused ultrasound modeling.\cite{ pasquinelli2020transducer, montanaro2021impact} The original hard sources (imposed pressure; sinusoidal with rise-time, or user-defined transient profiles) were extended for the purpose of this work by soft sources (cosine function to avoid slowly decaying low frequency components).  

The present benchmarks were simulated using isotropic voxel meshes (24 voxels per wavelength, 0.125 mm resolution) over the prescribed simulation domain padded with 96 layers of inhomogeneous PML, with a time step chosen to satisfy the CFL stability criterion (0.026 us or 76.9 PPP with bone, 0.048 us or 41.7 PPP without). 50 periods were simulated for benchmarks 1-7 (561 $\times$ 561 $\times$ 961 voxels), while 200 periods were simulated for benchhmark 8 (1521 $\times$ 1361 $\times$ 1801) and benchmark 9 (1473 $\times$ 1793 $\times$ 1697). To facilitate comparison, voxeling was offset by half a cell compared to the defined transducer surface, such that transducer grid points correspond to voxel cell centers and material interfaces to voxel faces. 

\subsection{STRIDE}

Stride solves the second-order, isotropic, linear acoustic wave equation using a FDTD approximation over a rectangular Cartesian grid.\cite{cueto2021stride} Spatial derivatives are calculated using a 10th-order finite-difference approximation, while time integration is performed using a 4th-order time-stepping scheme optimised for increased stability.\cite{amundsen2017time} Acoustic waves at the boundaries are absorbed using either a sponge absorbing boundary\cite{yao2018effective} or a complex frequency-shifted PML.\cite{gao2015unsplit}  Sources are introduced as free-field monopoles, which can be defined at locations both on and off the grid.\cite{hicks2002arbitrary}

Benchmarks 1 to 7 were computed using 24 PPW and 120 PPP, resulting in a grid size of 1061 $\times$ 661 $\times$ 661 including absorbing boundaries. Benchmarks 8 and 9 were computed using 18 PPW and 90 PPP, with grid sizes of 1451 $\times$ 1121 $\times$ 1241 and 1373 $\times$ 1445 $\times$ 1205, respectively. A complex frequency-shifted PML was used as the absorbing boundary for all benchmarks. Computed results were resampled onto the comparison mesh using linear interpolation.

\section{\label{sec_results}Benchmark Results}

\subsection{Field characteristics}

\begin{figure*}
\includegraphics[width=\textwidth]{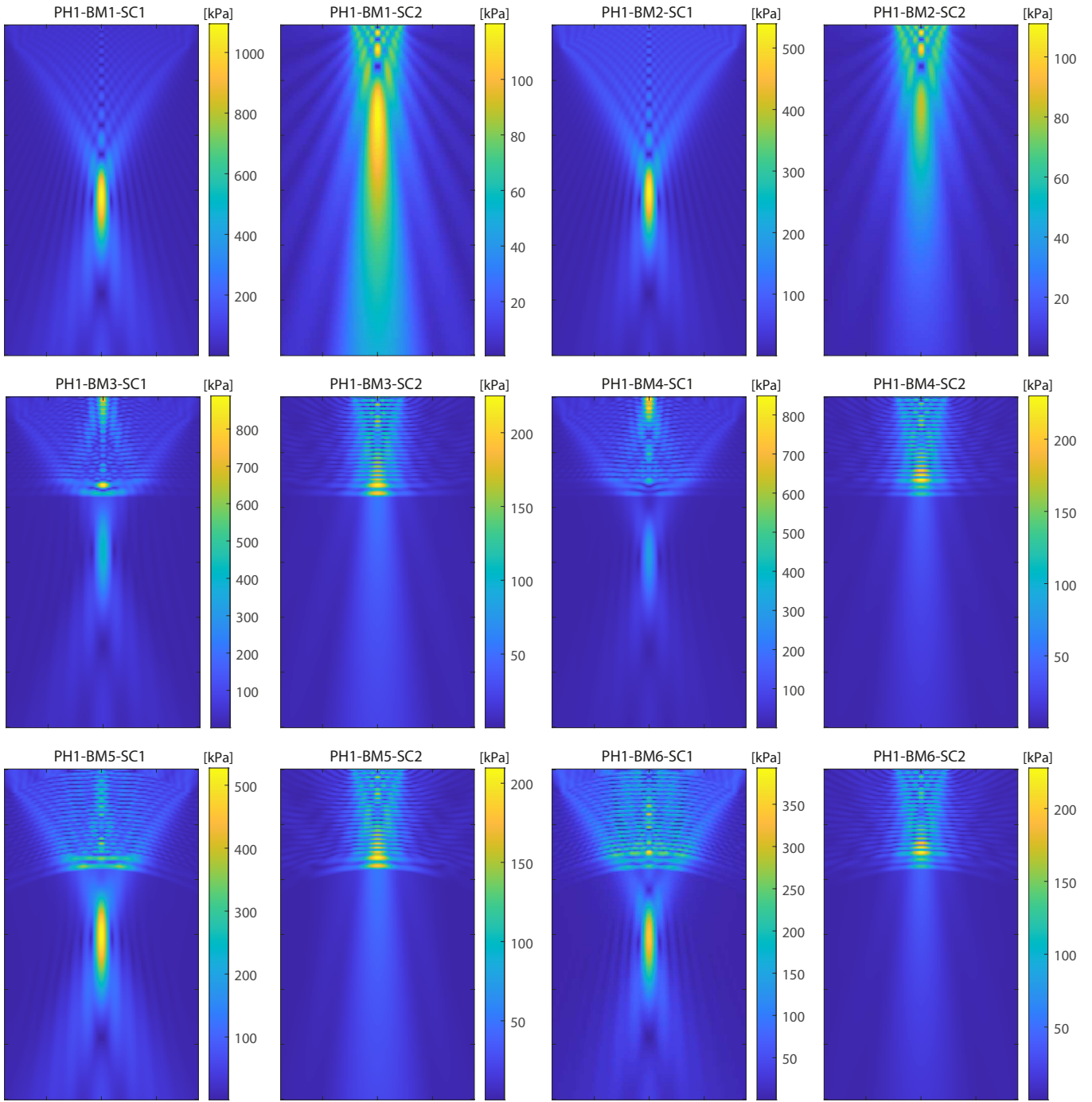}
\caption{\label{fig_example_fields_bm1to5}Pressure amplitudes computed using KWAVE for benchmarks 1 to 6 showing x-y slices through the central z-plane for a comparison domain of 120 mm (axial) by 70 mm (lateral). }
\raggedright
\end{figure*}

\begin{figure*}
\includegraphics[width=\textwidth]{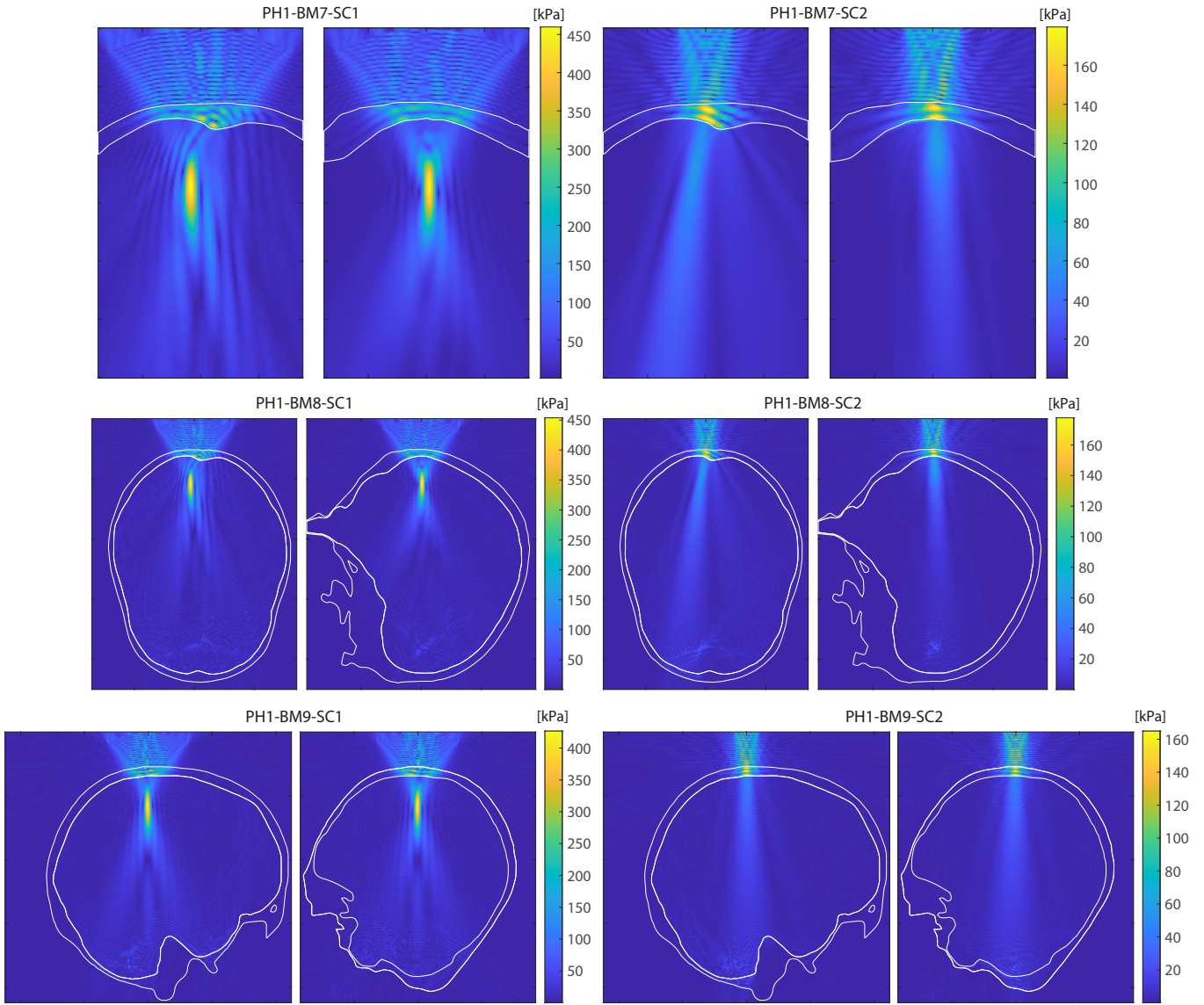}
\caption{\label{fig_example_fields_bm6to8}Pressure amplitudes computed using KWAVE for benchmarks 7 to 9 showing x-y (left) and x-z (right) slices through the location of the peak pressure. The approximate location of the skull is shown with the white overlay. The size of the comparison domain for each benchmark is given in Table \ref{tab_benchmarks}. }
\raggedright
\end{figure*}

Representative simulation results for all benchmarks are given in Figs.\ \ref{fig_example_fields_bm1to5} and \ref{fig_example_fields_bm6to8}. These illustrate the pressure amplitudes over the comparison domains given in Table \ref{tab_benchmarks}. The beam shapes for benchmark 1 and 2 are characteristic of focused bowl and unfocused piston transducers. The introduction of a flat skull bone with a single layer (benchmark 3) or multiple layers (benchmark 4) causes a drop in the focal pressure. Hot-spots (localized regions of increased pressure) are  introduced on the skull surface, and the reflected waves generate a complex interference pattern between the transducer and the skull. For the focused bowl transducer (\verb=PH1-BM3-SC1= and \verb=PH1-BM4-SC1=), the reflected waves also generate a secondary focus near the rear surface of the transducer. When a curved skull is used (benchmarks 5 and 6), the hot-spots and secondary focus are reduced. For all benchmarks with the piston transducer, a distinct last-axial maximum is no longer present after the introduction of the skull. Instead, the spatial peak pressure is typically either inside or immediately adjacent to the skull bone, and the acoustic beam gradually diverges after the skull surface. The introduction of a more complex skull geometry in benchmarks 7 to 9 generates additional features in the pressure fields. For benchmarks 7 and 8, the internal occipital protuberance of the skull bone causes a noticeable deflection of the acoustic beam. The use of the whole skull for benchmarks 8 and 9 also introduces small amplitude reflections from the opposite skull surface (e.g., see \verb=PH1-BM8-SC2= in Fig.\ \ref{fig_example_fields_bm6to8}).

\subsection{Difference metrics}

\begin{figure*}
\includegraphics[width=\textwidth]{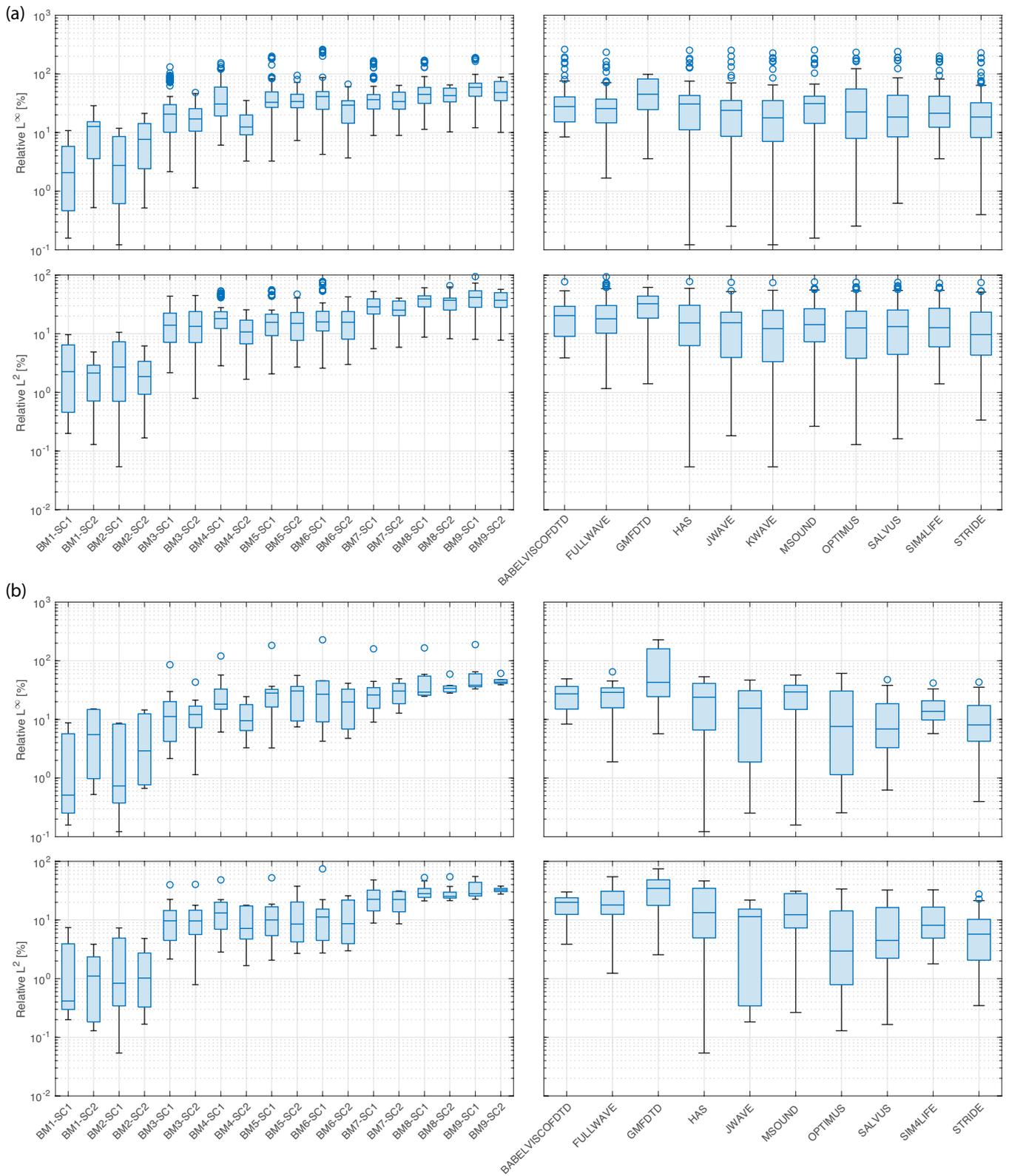}
\caption{\label{fig_intercomparison_linfl2}Summary of relative $L^\infty$ and $L^2$ difference metrics computed across the entire field taken from the exit plane of the transducer. (a) Cross comparison (all codes compared with all codes). (b) Comparison with KWAVE.}
\raggedright
\end{figure*}

\begin{figure*}
\includegraphics[width=\textwidth]{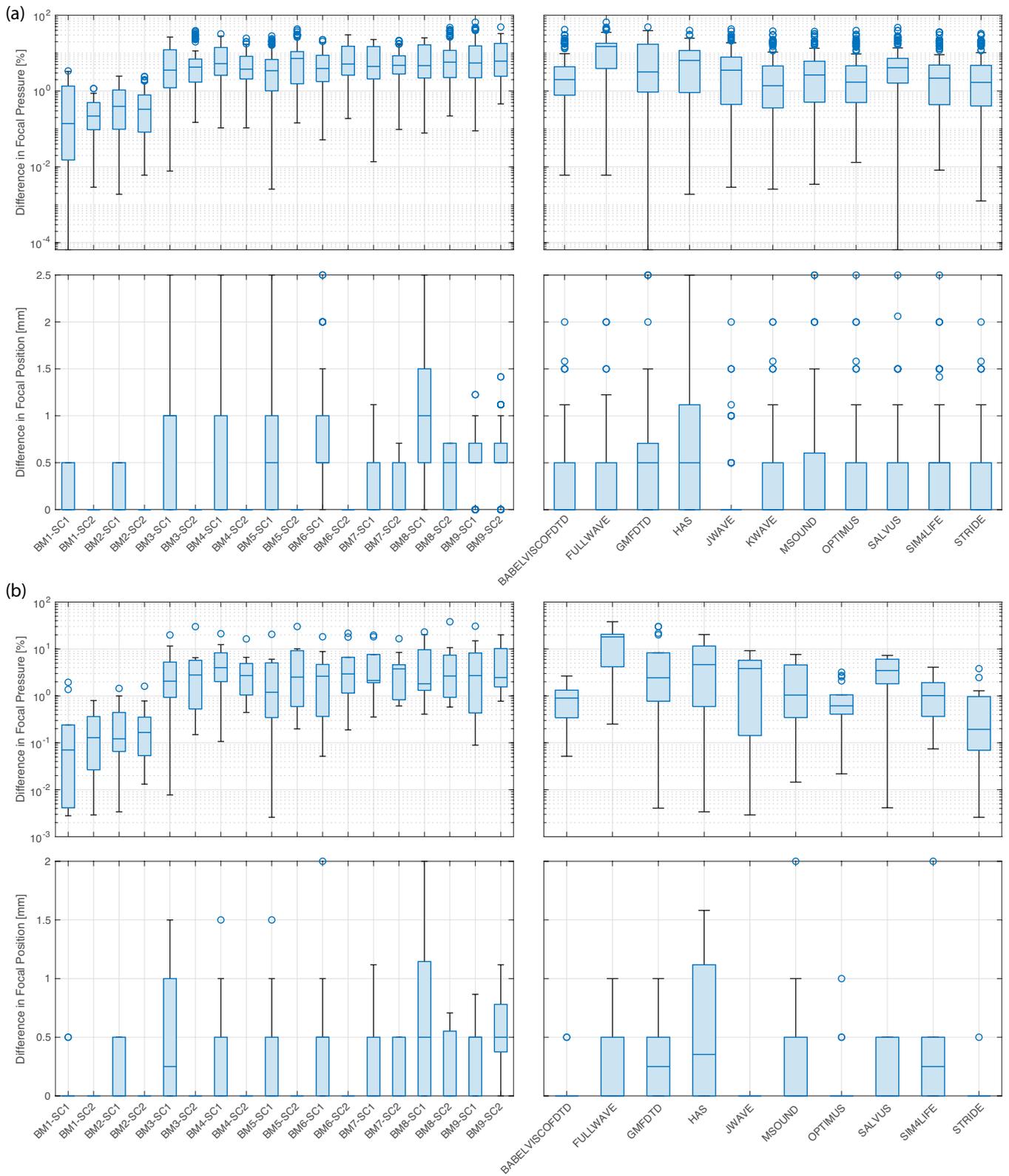}
\caption{\label{fig_intercomparison_focalpressure}Summary of focal (peak) pressure and focal position metrics. (a) Cross comparison (all codes compared with all codes). (b) Comparison with KWAVE.}
\raggedright
\end{figure*}

\begin{figure*}
\includegraphics[width=\textwidth]{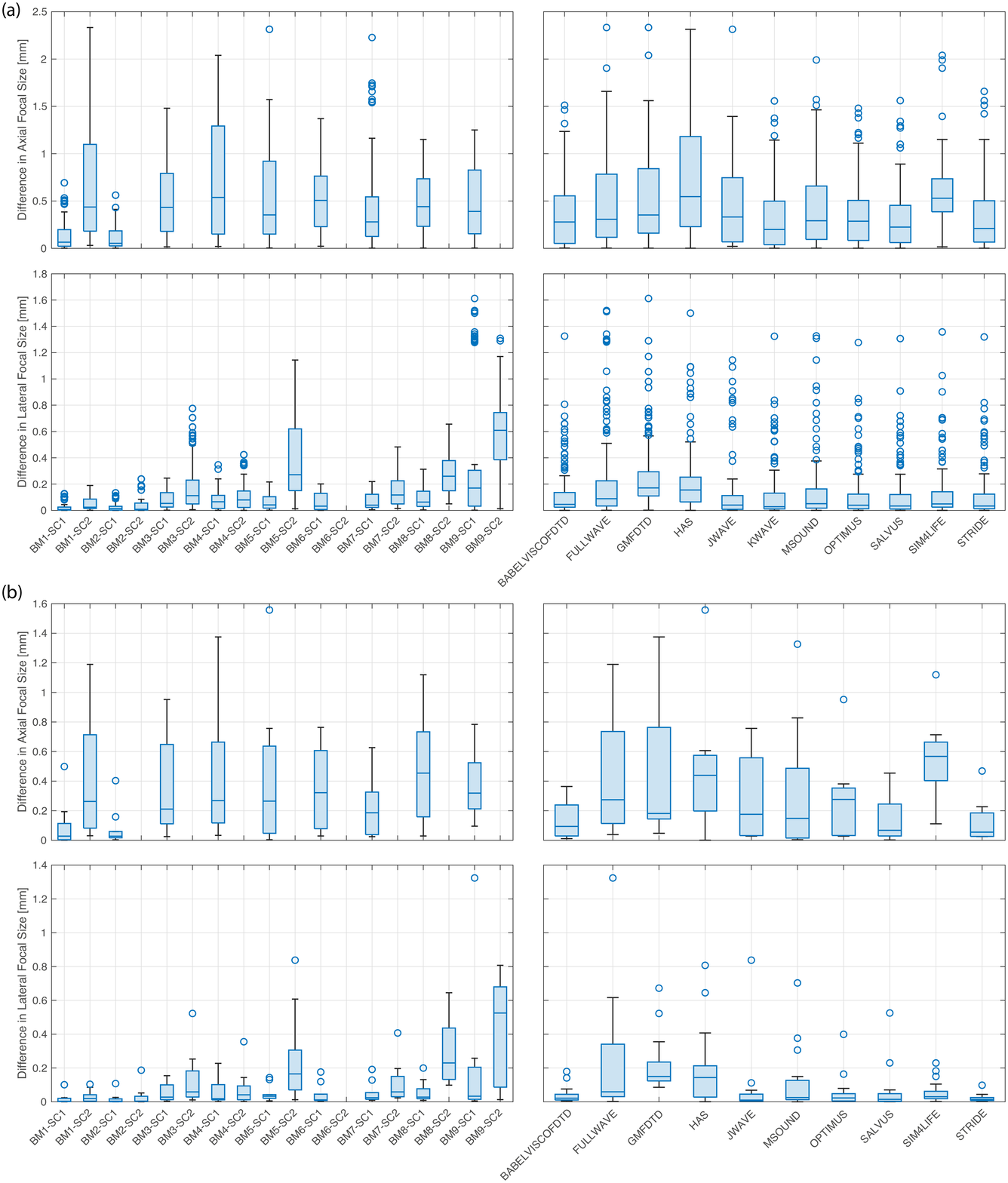}
\caption{\label{fig_intercomparison_focalsize}Summary of axial and lateral focal size metrics. Note, axial focal size was not computed for benchmarks 3 to 9 when using the plane piston source (SC2), as the field in this case did not have an axial maximum in the post-skull region. (a) Cross comparison (all codes compared with all codes). (b) Comparison with KWAVE.}
\raggedright
\end{figure*}

Aggregated difference metrics are given in Figs.\ \ref{fig_intercomparison_linfl2}-\ref{fig_intercomparison_focalsize}. These were calculated by comparing each model with every other model in a cross-comparison, and then computing the metrics described in Sec.\ \ref{sec_metrics}. The box plots (generated using \verb=boxchart= in MATLAB) illustrate the minimum, maximum, median, and first and third quartiles, along with any outliers. The same metrics were also computed for each model and benchmark using \verb=KWAVE= as a reference. This reference was used due to the very high spatial and temporal sampling possible for the \verb=KWAVE= simulations, particularly for benchmarks 1 to 6 which allowed an axisymmetric formulation to be used. Field plots, axial and lateral profiles, difference plots, and summary tables against \verb=FOCUS= (for benchmarks 1 and 2) and \verb=KWAVE= (for benchmarks 1 to 9) for each model are given in the Supplementary Material. These outputs are grouped both by benchmark and by model for ease of reference. Note, the simulation results and the comparison codes are freely available,\cite{results,code} thus it is straightforward to generate other comparisons as required, or add new modeling results to the intercomparison.

Figure \ref{fig_intercomparison_linfl2} gives a summary of the $L^\infty$ and $L^2$ intercomparison metrics computed across the comparison domains outlined in Table \ref{tab_benchmarks}. Results are presented for each benchmark (summarizing the cross-comparison results across all codes) and for each code (summarizing the cross-comparison results across all benchmarks). For benchmark 1 and 2 (water and water with artificial absorption), the level of agreement is very high. For the bowl transducer, seven models have $L^\infty$ values of less than 1\% when compared to FOCUS, and all values are less than 10\% (see Supplementary Material). For the piston transducer, the simulations are slightly less accurate. Four models have $L^\infty$ values of less than 1\% when compared to FOCUS, and the maximum $L^\infty$ value against FOCUS is 15\%. Examining the difference plots (see Supplementary Material), the largest differences are in the complex near-field pattern close to the transducer surface where the pressure varies rapidly.

For benchmarks 3 to 9, the $L^\infty$ and $L^2$ metrics both increase noticeably, with median values for the cross-comparison between 10\% and 100\% (Fig.\ \ref{fig_intercomparison_linfl2}(a)). There is still close agreement between some models, for example, three models have median $L^\infty$ values less than 10\% across all benchmarks when compared to \verb=KWAVE= (see Fig.\ \ref{fig_intercomparison_linfl2}(b)). However, in general, the differences are larger than those found for benchmarks 1 and 2. Examining the difference plots (see Supplementary Material), the largest variations are in the region between the transducer and skull bone. These arise due to a combination of errors in modeling the near-field of the transducer, even in free-field (described above), along with errors in modeling the reflection from the bone and soft-tissue surfaces (e.g., due to errors in the positions of the interfaces, and amplitude and phase errors in the reflected waves).

Overall, the $L^\infty$ and $L^2$ intercomparison metrics demonstrate that, on a pixel-by-pixel basis, there are often large variations between the model outputs. This is true, despite there being no uncertainty in the material parameters and transducer characteristics. This highlights the inherent uncertainties when using computational models for transcranial ultrasound simulation, which must be considered when interpreting model results. 

Figures \ref{fig_intercomparison_focalpressure} and \ref{fig_intercomparison_focalsize} give a summary of the intercomparison metrics for the focal position, size, and pressure. Despite the variations in the full-field error norms discussed above, there is very close agreement in the focal metrics. When compared by benchmark, the median values for the difference in focal pressure are all less than 10\% (see Fig.\ \ref{fig_intercomparison_focalpressure}(a)). Similarly, when compared by code, 10 out of 11 models have median differences less than 10\%. Differences of this level are on par with experimental repeatability and reproducibility measurements conducted using similar ultrasound transducers and a range of hydrophones.\cite{martin2019investigation} Compared to \verb=KWAVE=,
seven models have maximum differences in the focal pressure across all benchmarks of less than 10\%, and five models have median differences across all benchmarks on the order of 1\% or less (see Fig.\ \ref{fig_intercomparison_focalpressure}(b)). Considering the focal position, all values including outliers are within 2.5 mm (see Fig.\ \ref{fig_intercomparison_focalpressure}(a)), with median values for all benchmarks of 1 mm or less. Compared to \verb=KWAVE=, the median values for all models are less than 0.5 mm, with seven models having a median value of 0 mm (see Fig.\ \ref{fig_intercomparison_focalpressure}(b)).

Figure \ref{fig_intercomparison_focalsize} gives a summary of the intercomparison metrics for focal size. Note, as mentioned in Sec.\ \ref{sec_metrics}, the axial focal size for the piston transducer (\verb=SC2=) for benchmarks 3 to 9 is not calculated as there is no focus after propagation through the skull. For reference, in water (\verb=BM1=), the axial and lateral focal size for the focused bowl transducer is 26.2 and 4.1 mm, respectively, and the lateral focal size for the piston transducer at $x$ = 60 mm is 13.2 mm.  For all benchmarks, the median differences in the axial focal size for the focused bowl transducer are less than 0.6 mm (Fig.\ \ref{fig_intercomparison_focalsize}(a)), although there are a small number of outliers with differences up to 2.3 mm. The median differences in the lateral focal size for the focused bowl transducer for all benchmarks is 0.2 mm or less. Variations in the lateral focal size for the piston transducer are generally larger, noting the lateral focal size is also larger for this transducer. Similar results are evident for the comparison against \verb=KWAVE= (Fig.\ \ref{fig_intercomparison_focalsize}(b)).

Overall, there is very close agreement for all benchmarks in the characteristics of the focal pressure field after propagation through the skull bone. Larger differences are evident in the full-field metrics, dominated by differences in the field between the transducer and the skull. The most relevant metrics to compare, along with acceptable limits on the differences between models, depends strongly on the intended application of the computational results. For example, calculating phase delays, calculating the approximate position and size of the acoustic focus in the brain, and calculating the pressure in the skin and skull to subsequently estimate skull heating may each have different constraints and accuracy requirements. An analysis of these factors is beyond the scope of the current work. However, it is hoped that the benchmarks and computational results presented here may help to facilitate such investigations in the future.

\section{Summary}

A series of numerical benchmarks relevant to transcranial ultrasound simulation are presented, along with intercomparison results for 11 modeling tools used in the community. The intercomparison results show close agreement between the models, particularly for the position, size, and magnitude of the acoustic focus after propagating through the skull. When comparing each model with every other model in a cross comparison,  the median values for the difference in focal pressure and focal position are less than 10\% and 1 mm for all benchmarks. These results build confidence in the use of computational modeling to support transcranial ultrasound therapies. The benchmark definitions and associated data files, simulation results, and codes to compute the intercomparison metrics are all freely available.\cite{results,code} This allows the results to be replicated or further analysis to be conducted. Additional model results can also be easily added to the intercomparison, for example, to validate newly developed solvers. More generally, the intercomparison exercise provides a framework for creating benchmarks and performing model cross-comparisons. Further phases of the intercomparison exercise are currently under discussion, including benchmarks for elastic wave models, and model comparisons when using material parameters derived from CT images.

\section{Description of Supplementary material}
\verb=SuppPub1.xlsx= [URL to be inserted by AIP] gives an alternate table form of the model summaries given in Sec.\ \ref{sec_models}. \verb=SuppPub2.zip= [URL to be inserted by AIP] provides summaries of the comparison results (including metrics, field plots, axial profiles, and difference plots) for each model compared against \verb=FOCUS= (for benchmarks 1 and 2) and \verb=KWAVE= (for benchmarks 1 to 9). The \verb=.zip= file contains separate \verb=pdf= files for each model and for each benchmark, as a well as a summary of the cross-comparison metrics. The raw data files and MATLAB codes to process the results are also freely available.\cite{code,results}

\begin{acknowledgments}
The authors would like to thank the International Transcranial Ultrasonic Stimulation Safety and Standards (ITRUSST) consortium for providing the motivation and framework to conduct this work, and Robert McGough for helpful discussions regarding the use of FOCUS as a reference simulation. OB was supported by the Engineering and Physical Sciences Research Council (EPSRC) Centre for Doctoral Training in Neurotechnology, grant number EP/L016737/1. KBP and NL acknowledge the support of the National Institutes of Health (NIH), grant numbers R01 CA227687, NIH T32 EB009653, and NSF DGE 1656518. DC acknowledges support from the Focused Ultrasound Foundation, and the NIH grants R01 EB013433, R01 CA172787, R01 EB028316, R37 CA224141. CC acknowledges the support of the EPSRC, grant number EP/T51780X/1. PG and EvW acknowledge support of the EPSRC, grant number EP/P012434/1, and use of the UCL Myriad High Performance Computing Facility (Myriad@UCL), and associated support services. JJ was supported by Brno University of Technology under project number FIT-S-20-6309. YJ acknowledges the support of the NIH, grant number R01EB025205. PM and CB acknowledge support from the Swiss National Supercomputing Centre (CSCS) under project IDs s1040 and sm59. SP acknowledges the support of the Natural Sciences and Engineering Research Council of Canada and the Canada Foundation for Innovation. AP would like to acknowledge Academy of Finland projects 320166, 336119, 336799. AS and BT were supported by the EPSRC, grant number EP/S026371/1. AT was supported by the Lundbeck foundation, grant number R313-2019-622.
\end{acknowledgments}

\bibliography{references.bib}

\begin{thebibliography}{10}
\def\enquote#1,{``#1,''}
\def\enxquote#1{``#1''}
\expandafter\ifx\csname url\endcsname\relax
  \def\url#1{\texttt{#1}}\fi
\expandafter\ifx\csname urlprefix\endcsname\relax\def\urlprefix{URL }\fi
\providecommand{\bibinfo}[2]{#2}
\def\plainquote#1{``#1''}
\providecommand{\noopsort}[1]{}
\providecommand{\switchargs}[2]{#2#1}
\providecommand{\dourl}[1]{\href{http://#1}{\nolinkurl{#1}}}
  \def\eatspace #1{#1}

\bibitem{elias2016randomized}
\bibinfo{author}{W.~J. Elias}, \bibinfo{author}{N.~Lipsman},
  \bibinfo{author}{W.~G. Ondo}, \bibinfo{author}{P.~Ghanouni},
  \bibinfo{author}{Y.~G. Kim}, \bibinfo{author}{W.~Lee},
  \bibinfo{author}{M.~Schwartz}, \bibinfo{author}{K.~Hynynen},
  \bibinfo{author}{A.~M. Lozano}, \bibinfo{author}{B.~B. Shah}, \emph{et~al.},
  \enquote{\bibinfo{title}{A randomized trial of focused ultrasound thalamotomy
  for essential tremor}},  \bibinfo{journal}{New England Journal of Medicine}
  \textbf{375}(8), \bibinfo{pages}{730--739} (\bibinfo{year}{2016}).

\bibitem{abrahao2019first}
\bibinfo{author}{A.~Abrahao}, \bibinfo{author}{Y.~Meng},
  \bibinfo{author}{M.~Llinas}, \bibinfo{author}{Y.~Huang},
  \bibinfo{author}{C.~Hamani}, \bibinfo{author}{T.~Mainprize},
  \bibinfo{author}{I.~Aubert}, \bibinfo{author}{C.~Heyn},
  \bibinfo{author}{S.~E. Black}, \bibinfo{author}{K.~Hynynen},
  \bibinfo{author}{N.~Lipsman}, and \bibinfo{author}{L.~Zinman},
  \enquote{\bibinfo{title}{First-in-human trial of blood--brain barrier opening
  in amyotrophic lateral sclerosis using {MR}-guided focused ultrasound}},
  \bibinfo{journal}{Nature Communications} \textbf{10}(1),
  \bibinfo{pages}{1--9} (\bibinfo{year}{2019}).

\bibitem{legon2014transcranial}
\bibinfo{author}{W.~Legon}, \bibinfo{author}{T.~F. Sato},
  \bibinfo{author}{A.~Opitz}, \bibinfo{author}{J.~Mueller},
  \bibinfo{author}{A.~Barbour}, \bibinfo{author}{A.~Williams}, and
  \bibinfo{author}{W.~J. Tyler}, \enquote{\bibinfo{title}{Transcranial focused
  ultrasound modulates the activity of primary somatosensory cortex in
  humans}},  \bibinfo{journal}{Nature Neuroscience} \textbf{17}(2),
  \bibinfo{pages}{322--329} (\bibinfo{year}{2014}).

\bibitem{hynynen1998demonstration}
\bibinfo{author}{K.~Hynynen} and \bibinfo{author}{F.~A. Jolesz},
  \enquote{\bibinfo{title}{Demonstration of potential noninvasive ultrasound
  brain therapy through an intact skull}},  \bibinfo{journal}{Ultrasound in
  Medicine \& Biology} \textbf{24}(2), \bibinfo{pages}{275--283}
  (\bibinfo{year}{1998}).

\bibitem{bouchoux2012experimental}
\bibinfo{author}{G.~Bouchoux}, \bibinfo{author}{K.~B. Bader},
  \bibinfo{author}{J.~J. Korfhagen}, \bibinfo{author}{J.~L. Raymond},
  \bibinfo{author}{R.~Shivashankar}, \bibinfo{author}{T.~A. Abruzzo}, and
  \bibinfo{author}{C.~K. Holland}, \enquote{\bibinfo{title}{Experimental
  validation of a finite-difference model for the prediction of transcranial
  ultrasound fields based on {CT} images}},  \bibinfo{journal}{Physics in
  Medicine \& Biology} \textbf{57}(23), \bibinfo{pages}{8005}
  (\bibinfo{year}{2012}).

\bibitem{marquet2009non}
\bibinfo{author}{F.~Marquet}, \bibinfo{author}{M.~Pernot},
  \bibinfo{author}{J.-F. Aubry}, \bibinfo{author}{G.~Montaldo},
  \bibinfo{author}{L.~Marsac}, \bibinfo{author}{M.~Tanter}, and
  \bibinfo{author}{M.~Fink}, \enquote{\bibinfo{title}{Non-invasive transcranial
  ultrasound therapy based on a {3D} {CT} scan: protocol validation and in
  vitro results}},  \bibinfo{journal}{Physics in Medicine \& Biology}
  \textbf{54}(9), \bibinfo{pages}{2597} (\bibinfo{year}{2009}).

\bibitem{dallapiazza2017noninvasive}
\bibinfo{author}{R.~F. Dallapiazza}, \bibinfo{author}{K.~F. Timbie},
  \bibinfo{author}{S.~Holmberg}, \bibinfo{author}{J.~Gatesman},
  \bibinfo{author}{M.~B. Lopes}, \bibinfo{author}{R.~J. Price},
  \bibinfo{author}{G.~W. Miller}, and \bibinfo{author}{W.~J. Elias},
  \enquote{\bibinfo{title}{Noninvasive neuromodulation and thalamic mapping
  with low-intensity focused ultrasound}},  \bibinfo{journal}{Journal of
  Neurosurgery} \textbf{128}(3), \bibinfo{pages}{875--884}
  (\bibinfo{year}{2017}).

\bibitem{code}
\bibinfo{author}{B.~Treeby}, \enxquote{\bibinfo{title}{Benchmark problems for
  transcranial ultrasound simulation: Intercomparison library}} ,
  \bibinfo{howpublished}{(v1.0) [Code] GitHub} (\bibinfo{year}{2022}),
  \dourl{https://github.com/ucl-bug/transcranial-ultrasound-benchmarks}.

\bibitem{results}
\bibinfo{author}{J.-F. Aubry}, \bibinfo{author}{O.~Bates},
  \bibinfo{author}{C.~Boehm}, \bibinfo{author}{K.~{Butts Pauly}},
  \bibinfo{author}{D.~Christensen}, \bibinfo{author}{C.~Cueto},
  \bibinfo{author}{P.~Gelat}, \bibinfo{author}{L.~Guasch},
  \bibinfo{author}{J.~Jaros}, \bibinfo{author}{Y.~Jing},
  \bibinfo{author}{R.~Jones}, \bibinfo{author}{N.~Li},
  \bibinfo{author}{P.~Marty}, \bibinfo{author}{H.~Montanaro},
  \bibinfo{author}{E.~Neufeld}, \bibinfo{author}{S.~Pichardo},
  \bibinfo{author}{G.~Pinton}, \bibinfo{author}{A.~Pulkkinen},
  \bibinfo{author}{A.~Stanziola}, \bibinfo{author}{A.~Thielscher},
  \bibinfo{author}{B.~Treeby}, and \bibinfo{author}{E.~{van 't Wout}},
  \enxquote{\bibinfo{title}{{Benchmark problems for transcranial ultrasound
  simulation: Datasets for intercomparison of compressional wave models}}} ,
  \bibinfo{howpublished}{(v1.0) [Dataset], Zenodo} (\bibinfo{year}{2022}),
  \dodoi{{https://doi.org/10.5281/zenodo.6020543}}.

\bibitem{clement2004enhanced}
\bibinfo{author}{G.~T. Clement}, \bibinfo{author}{P.~J. White}, and
  \bibinfo{author}{K.~Hynynen}, \enquote{\bibinfo{title}{Enhanced ultrasound
  transmission through the human skull using shear mode conversion}},
  \bibinfo{journal}{The Journal of the Acoustical Society of America}
  \textbf{115}(3), \bibinfo{pages}{1356--1364} (\bibinfo{year}{2004}).

\bibitem{younan2013influence}
\bibinfo{author}{Y.~Younan}, \bibinfo{author}{T.~Deffieux},
  \bibinfo{author}{B.~Larrat}, \bibinfo{author}{M.~Fink},
  \bibinfo{author}{M.~Tanter}, and \bibinfo{author}{J.-F. Aubry},
  \enquote{\bibinfo{title}{Influence of the pressure field distribution in
  transcranial ultrasonic neurostimulation}},  \bibinfo{journal}{Medical
  Physics} \textbf{40}(8), \bibinfo{pages}{082902} (\bibinfo{year}{2013}).

\bibitem{robertson2017accurate}
\bibinfo{author}{J.~L. Robertson}, \bibinfo{author}{B.~T. Cox},
  \bibinfo{author}{J.~Jaros}, and \bibinfo{author}{B.~E. Treeby},
  \enquote{\bibinfo{title}{Accurate simulation of transcranial ultrasound
  propagation for ultrasonic neuromodulation and stimulation}},
  \bibinfo{journal}{The Journal of the Acoustical Society of America}
  \textbf{141}(3), \bibinfo{pages}{1726--1738} (\bibinfo{year}{2017}).

\bibitem{o1949theory}
\bibinfo{author}{H.~O'Neil}, \enquote{\bibinfo{title}{Theory of focusing
  radiators}},  \bibinfo{journal}{The Journal of the Acoustical Society of
  America} \textbf{21}(5), \bibinfo{pages}{516--526} (\bibinfo{year}{1949}).

\bibitem{fry1978acoustical}
\bibinfo{author}{F.~J. Fry} and \bibinfo{author}{J.~E. Barger},
  \enquote{\bibinfo{title}{Acoustical properties of the human skull}},
  \bibinfo{journal}{The Journal of the Acoustical Society of America}
  \textbf{63}(5), \bibinfo{pages}{1576--1590} (\bibinfo{year}{1978}).

\bibitem{mast2000empirical}
\bibinfo{author}{T.~D. Mast}, \enquote{\bibinfo{title}{Empirical relationships
  between acoustic parameters in human soft tissues}},
  \bibinfo{journal}{Acoustics Research Letters Online} \textbf{1}(2),
  \bibinfo{pages}{37--42} (\bibinfo{year}{2000}).

\bibitem{clement2002correlation}
\bibinfo{author}{G.~Clement} and \bibinfo{author}{K.~Hynynen},
  \enquote{\bibinfo{title}{Correlation of ultrasound phase with physical skull
  properties}},  \bibinfo{journal}{Ultrasound in Medicine \& Biology}
  \textbf{28}(5), \bibinfo{pages}{617--624} (\bibinfo{year}{2002}).

\bibitem{pichardo2010multi}
\bibinfo{author}{S.~Pichardo}, \bibinfo{author}{V.~W. Sin}, and
  \bibinfo{author}{K.~Hynynen}, \enquote{\bibinfo{title}{Multi-frequency
  characterization of the speed of sound and attenuation coefficient for
  longitudinal transmission of freshly excised human skulls}},
  \bibinfo{journal}{Physics in Medicine \& Biology} \textbf{56}(1),
  \bibinfo{pages}{219} (\bibinfo{year}{2010}).

\bibitem{pinton2012attenuation}
\bibinfo{author}{G.~Pinton}, \bibinfo{author}{J.-F. Aubry},
  \bibinfo{author}{E.~Bossy}, \bibinfo{author}{M.~Muller},
  \bibinfo{author}{M.~Pernot}, and \bibinfo{author}{M.~Tanter},
  \enquote{\bibinfo{title}{Attenuation, scattering, and absorption of
  ultrasound in the skull bone}},  \bibinfo{journal}{Medical Physics}
  \textbf{39}(1), \bibinfo{pages}{299--307} (\bibinfo{year}{2012}).

\bibitem{pichardo2017viscoelastic}
\bibinfo{author}{S.~Pichardo}, \bibinfo{author}{C.~Moreno-Hern{\'a}ndez},
  \bibinfo{author}{R.~A. Drainville}, \bibinfo{author}{V.~Sin},
  \bibinfo{author}{L.~Curiel}, and \bibinfo{author}{K.~Hynynen},
  \enquote{\bibinfo{title}{A viscoelastic model for the prediction of
  transcranial ultrasound propagation: application for the estimation of shear
  acoustic properties in the human skull}},  \bibinfo{journal}{Physics in
  Medicine \& Biology} \textbf{62}(17), \bibinfo{pages}{6938}
  (\bibinfo{year}{2017}).

\bibitem{white2006longitudinal}
\bibinfo{author}{P.~J. White}, \bibinfo{author}{G.~T. Clement}, and
  \bibinfo{author}{K.~Hynynen}, \enquote{\bibinfo{title}{Longitudinal and shear
  mode ultrasound propagation in human skull bone}},
  \bibinfo{journal}{Ultrasound in Medicine \& Biology} \textbf{32}(7),
  \bibinfo{pages}{1085--1096} (\bibinfo{year}{2006}).

\bibitem{webb2020acoustic}
\bibinfo{author}{T.~D. Webb}, \bibinfo{author}{S.~A. Leung},
  \bibinfo{author}{P.~Ghanouni}, \bibinfo{author}{J.~J. Dahl},
  \bibinfo{author}{N.~J. Pelc}, and \bibinfo{author}{K.~B. Pauly},
  \enquote{\bibinfo{title}{Acoustic attenuation: Multifrequency measurement and
  relationship to {CT} and {MR} imaging}},  \bibinfo{journal}{IEEE Transactions
  on Ultrasonics, Ferroelectrics, and Frequency Control} \textbf{68}(5),
  \bibinfo{pages}{1532--1545} (\bibinfo{year}{2020}).

\bibitem{mcgough2004efficient}
\bibinfo{author}{R.~J. McGough}, \bibinfo{author}{T.~V. Samulski}, and
  \bibinfo{author}{J.~F. Kelly}, \enquote{\bibinfo{title}{An efficient grid
  sectoring method for calculations of the near-field pressure generated by a
  circular piston}},  \bibinfo{journal}{The Journal of the Acoustical Society
  of America} \textbf{115}(5), \bibinfo{pages}{1942--1954}
  (\bibinfo{year}{2004}).

\bibitem{chen20082d}
\bibinfo{author}{D.~Chen} and \bibinfo{author}{R.~J. McGough},
  \enquote{\bibinfo{title}{A {2D} fast near-field method for calculating
  near-field pressures generated by apodized rectangular pistons}},
  \bibinfo{journal}{The Journal of the Acoustical Society of America}
  \textbf{124}(3), \bibinfo{pages}{1526--1537} (\bibinfo{year}{2008}).

\bibitem{kelly2009transient}
\bibinfo{author}{J.~F. Kelly} and \bibinfo{author}{R.~J. McGough},
  \enquote{\bibinfo{title}{Transient fields generated by spherical shells in
  viscous media}}, in \emph{\bibinfo{booktitle}{AIP Conference Proceedings}},
  \bibinfo{organization}{American Institute of Physics} (\bibinfo{year}{2009}),
  Vol. \bibinfo{volume}{1113}, pp. \bibinfo{pages}{210--214}.

\bibitem{alexander2019structural}
\bibinfo{author}{S.~L. Alexander}, \bibinfo{author}{K.~Rafaels},
  \bibinfo{author}{C.~A. Gunnarsson}, and \bibinfo{author}{T.~Weerasooriya},
  \enquote{\bibinfo{title}{Structural analysis of the frontal and parietal
  bones of the human skull}},  \bibinfo{journal}{Journal of the Mechanical
  Behavior of Biomedical Materials} \textbf{90}, \bibinfo{pages}{689--701}
  (\bibinfo{year}{2019}).

\bibitem{hori1972thickness}
\bibinfo{author}{H.~Hori}, \bibinfo{author}{G.~Moretti},
  \bibinfo{author}{A.~Rebora}, and \bibinfo{author}{F.~Crovato},
  \enquote{\bibinfo{title}{The thickness of human scalp: normal and bald}},
  \bibinfo{journal}{Journal of Investigative Dermatology} \textbf{58}(6),
  \bibinfo{pages}{396--399} (\bibinfo{year}{1972}).

\bibitem{fonov2009unbiased}
\bibinfo{author}{V.~S. Fonov}, \bibinfo{author}{A.~C. Evans},
  \bibinfo{author}{R.~C. McKinstry}, \bibinfo{author}{C.~Almli}, and
  \bibinfo{author}{D.~Collins}, \enquote{\bibinfo{title}{Unbiased nonlinear
  average age-appropriate brain templates from birth to adulthood}},
  \bibinfo{journal}{NeuroImage} \textbf{47}, \bibinfo{pages}{S102}
  (\bibinfo{year}{2009}).

\bibitem{fonov2011unbiased}
\bibinfo{author}{V.~Fonov}, \bibinfo{author}{A.~C. Evans},
  \bibinfo{author}{K.~Botteron}, \bibinfo{author}{C.~R. Almli},
  \bibinfo{author}{R.~C. McKinstry}, \bibinfo{author}{D.~L. Collins}, and
  \bibinfo{author}{{Brain Development Cooperative Group}},
  \enquote{\bibinfo{title}{Unbiased average age-appropriate atlases for
  pediatric studies}},  \bibinfo{journal}{NeuroImage} \textbf{54}(1),
  \bibinfo{pages}{313--327} (\bibinfo{year}{2011}).

\bibitem{nielsen2018automatic}
\bibinfo{author}{J.~D. Nielsen}, \bibinfo{author}{K.~H. Madsen},
  \bibinfo{author}{O.~Puonti}, \bibinfo{author}{H.~R. Siebner},
  \bibinfo{author}{C.~Bauer}, \bibinfo{author}{C.~G. Madsen},
  \bibinfo{author}{G.~B. Saturnino}, and \bibinfo{author}{A.~Thielscher},
  \enquote{\bibinfo{title}{Automatic skull segmentation from {MR} images for
  realistic volume conductor models of the head: Assessment of the
  state-of-the-art}},  \bibinfo{journal}{NeuroImage} \textbf{174},
  \bibinfo{pages}{587--598} (\bibinfo{year}{2018}).

\bibitem{iso2mesh}
\bibinfo{author}{Q.~Fang} and \bibinfo{author}{D.~A. Boas},
  \enquote{\bibinfo{title}{Tetrahedral mesh generation from volumetric binary
  and grayscale images}}, in \emph{\bibinfo{booktitle}{2009 IEEE International
  Symposium on Biomedical Imaging: From Nano to Macro}},
  \bibinfo{organization}{Ieee} (\bibinfo{year}{2009}), pp.
  \bibinfo{pages}{1142--1145}.

\bibitem{fang2009tetrahedral}
\bibinfo{author}{Q.~Fang} and \bibinfo{author}{D.~A. Boas},
  \enquote{\bibinfo{title}{Tetrahedral mesh generation from volumetric binary
  and grayscale images}}, in \emph{\bibinfo{booktitle}{IEEE International
  Symposium on Biomedical Imaging: From Nano to Macro}},
  \bibinfo{organization}{Ieee} (\bibinfo{year}{2009}), pp.
  \bibinfo{pages}{1142--1145}.

\bibitem{virieux1986p}
\bibinfo{author}{J.~Virieux}, \enquote{\bibinfo{title}{{P-SV} wave propagation
  in heterogeneous media: Velocity-stress finite-difference method}},
  \bibinfo{journal}{Geophysics} \textbf{51}(4), \bibinfo{pages}{889--901}
  (\bibinfo{year}{1986}).

\bibitem{levander1988fourth}
\bibinfo{author}{A.~R. Levander}, \enquote{\bibinfo{title}{Fourth-order
  finite-difference {P-SV} seismograms}},  \bibinfo{journal}{Geophysics}
  \textbf{53}(11), \bibinfo{pages}{1425--1436} (\bibinfo{year}{1988}).

\bibitem{bohlen2002parallel}
\bibinfo{author}{T.~Bohlen}, \enquote{\bibinfo{title}{Parallel {3-D}
  viscoelastic finite difference seismic modelling}},
  \bibinfo{journal}{Computers \& Geosciences} \textbf{28}(8),
  \bibinfo{pages}{887--899} (\bibinfo{year}{2002}).

\bibitem{blanch1995modeling}
\bibinfo{author}{J.~O. Blanch}, \bibinfo{author}{J.~O. Robertsson}, and
  \bibinfo{author}{W.~W. Symes}, \enquote{\bibinfo{title}{Modeling of a
  constant {Q}: Methodology and algorithm for an efficient and optimally
  inexpensive viscoelastic technique}},  \bibinfo{journal}{Geophysics}
  \textbf{60}(1), \bibinfo{pages}{176--184} (\bibinfo{year}{1995}).

\bibitem{moczo20023d}
\bibinfo{author}{P.~Moczo}, \bibinfo{author}{J.~Kristek},
  \bibinfo{author}{V.~Vavrycuk}, \bibinfo{author}{R.~J. Archuleta}, and
  \bibinfo{author}{L.~Halada}, \enquote{\bibinfo{title}{{3D} heterogeneous
  staggered-grid finite-difference modeling of seismic motion with volume
  harmonic and arithmetic averaging of elastic moduli and densities}},
  \bibinfo{journal}{Bulletin of the Seismological Society of America}
  \textbf{92}(8), \bibinfo{pages}{3042--3066} (\bibinfo{year}{2002}).

\bibitem{drainville2019superposition}
\bibinfo{author}{R.~A. Drainville}, \bibinfo{author}{L.~Curiel}, and
  \bibinfo{author}{S.~Pichardo}, \enquote{\bibinfo{title}{Superposition method
  for modelling boundaries between media in viscoelastic finite difference time
  domain simulations}},  \bibinfo{journal}{The Journal of the Acoustical
  Society of America} \textbf{146}(6), \bibinfo{pages}{4382--4401}
  (\bibinfo{year}{2019}).

\bibitem{collino2001application}
\bibinfo{author}{F.~Collino} and \bibinfo{author}{C.~Tsogka},
  \enquote{\bibinfo{title}{Application of the perfectly matched absorbing layer
  model to the linear elastodynamic problem in anisotropic heterogeneous
  media}},  \bibinfo{journal}{Geophysics} \textbf{66}(1),
  \bibinfo{pages}{294--307} (\bibinfo{year}{2001}).

\bibitem{pinton2009heterogeneous}
\bibinfo{author}{G.~F. Pinton}, \bibinfo{author}{J.~Dahl},
  \bibinfo{author}{S.~Rosenzweig}, and \bibinfo{author}{G.~E. Trahey},
  \enquote{\bibinfo{title}{A heterogeneous nonlinear attenuating full-wave
  model of ultrasound}},  \bibinfo{journal}{IEEE Transactions on Ultrasonics,
  Ferroelectrics, and Frequency Control} \textbf{56}(3),
  \bibinfo{pages}{474--488} (\bibinfo{year}{2009}).

\bibitem{pinton2021fullwave}
\bibinfo{author}{G.~Pinton}, \enquote{\bibinfo{title}{A fullwave model of the
  nonlinear wave equation with multiple relaxations and relaxing perfectly
  matched layers for high-order numerical finite-difference solutions}},
  \bibinfo{journal}{arXiv preprint arXiv:2106.11476}  (\bibinfo{year}{2021}).

\bibitem{courant1928partiellen}
\bibinfo{author}{R.~Courant}, \bibinfo{author}{K.~Friedrichs}, and
  \bibinfo{author}{H.~Lewy}, \enquote{\bibinfo{title}{{\"U}ber die partiellen
  differenzengleichungen der mathematischen physik}},
  \bibinfo{journal}{Mathematische Annalen} \textbf{100}(1),
  \bibinfo{pages}{32--74} (\bibinfo{year}{1928}).

\bibitem{pulkkinen2014numerical}
\bibinfo{author}{A.~Pulkkinen}, \bibinfo{author}{B.~Werner},
  \bibinfo{author}{E.~Martin}, and \bibinfo{author}{K.~Hynynen},
  \enquote{\bibinfo{title}{Numerical simulations of clinical focused ultrasound
  functional neurosurgery}},  \bibinfo{journal}{Physics in Medicine \& Biology}
  \textbf{59}(7), \bibinfo{pages}{1679} (\bibinfo{year}{2014}).

\bibitem{vyas2012ultrasound}
\bibinfo{author}{U.~Vyas} and \bibinfo{author}{D.~Christensen},
  \enquote{\bibinfo{title}{Ultrasound beam simulations in inhomogeneous tissue
  geometries using the hybrid angular spectrum method}},
  \bibinfo{journal}{IEEE Transactions on Ultrasonics, Ferroelectrics, and
  Frequency Control} \textbf{59}(6), \bibinfo{pages}{1093--1100}
  (\bibinfo{year}{2012}).

\bibitem{almquist2016rapid}
\bibinfo{author}{S.~Almquist}, \bibinfo{author}{D.~L. Parker}, and
  \bibinfo{author}{D.~A. Christensen}, \enquote{\bibinfo{title}{Rapid full-wave
  phase aberration correction method for transcranial high-intensity focused
  ultrasound therapies}},  \bibinfo{journal}{Journal of Therapeutic Ultrasound}
  \textbf{4}(1), \bibinfo{pages}{1--11} (\bibinfo{year}{2016}).

\bibitem{saad1986gmres}
\bibinfo{author}{Y.~Saad} and \bibinfo{author}{M.~H. Schultz},
  \enquote{\bibinfo{title}{{GMRES}: A generalized minimal residual algorithm
  for solving nonsymmetric linear systems}},  \bibinfo{journal}{SIAM Journal on
  Scientific and Statistical Computing} \textbf{7}(3),
  \bibinfo{pages}{856--869} (\bibinfo{year}{1986}).

\bibitem{bermudez2007optimal}
\bibinfo{author}{A.~Berm{\'u}dez}, \bibinfo{author}{L.~Hervella-Nieto},
  \bibinfo{author}{A.~Prieto}, and \bibinfo{author}{R.~Rodr{\i}},
  \enquote{\bibinfo{title}{An optimal perfectly matched layer with unbounded
  absorbing function for time-harmonic acoustic scattering problems}},
  \bibinfo{journal}{Journal of Computational Physics} \textbf{223}(2),
  \bibinfo{pages}{469--488} (\bibinfo{year}{2007}).

\bibitem{wise2019representing}
\bibinfo{author}{E.~S. Wise}, \bibinfo{author}{B.~Cox},
  \bibinfo{author}{J.~Jaros}, and \bibinfo{author}{B.~E. Treeby},
  \enquote{\bibinfo{title}{Representing arbitrary acoustic source and sensor
  distributions in {Fourier} collocation methods}},  \bibinfo{journal}{The
  Journal of the Acoustical Society of America} \textbf{146}(1),
  \bibinfo{pages}{278--288} (\bibinfo{year}{2019}).

\bibitem{stanziola2021research}
\bibinfo{author}{A.~Stanziola}, \bibinfo{author}{S.~R. Arridge},
  \bibinfo{author}{B.~T. Cox}, and \bibinfo{author}{B.~E. Treeby},
  \enquote{\bibinfo{title}{A research framework for writing differentiable
  {PDE} discretizations in {JAX}}},  \bibinfo{journal}{arXiv preprint
  arXiv:2111.05218}  (\bibinfo{year}{2021}).

\bibitem{jax2018github}
\bibinfo{author}{J.~Bradbury}, \bibinfo{author}{R.~Frostig},
  \bibinfo{author}{P.~Hawkins}, \bibinfo{author}{M.~J. Johnson},
  \bibinfo{author}{C.~Leary}, \bibinfo{author}{D.~Maclaurin},
  \bibinfo{author}{G.~Necula}, \bibinfo{author}{A.~Paszke},
  \bibinfo{author}{J.~Vander{P}las}, \bibinfo{author}{S.~Wanderman-{M}ilne},
  and \bibinfo{author}{Q.~Zhang}, \enxquote{\bibinfo{title}{{JAX}: composable
  transformations of {P}ython+{N}um{P}y programs}}  (\bibinfo{year}{2018}),
  \dourl{http://github.com/google/jax}.

\bibitem{treeby2010k}
\bibinfo{author}{B.~E. Treeby} and \bibinfo{author}{B.~T. Cox},
  \enquote{\bibinfo{title}{{k-Wave}: {MATLAB} toolbox for the simulation and
  reconstruction of photoacoustic wave fields}},  \bibinfo{journal}{Journal of
  Biomedical Optics} \textbf{15}(2), \bibinfo{pages}{021314}
  (\bibinfo{year}{2010}).

\bibitem{treeby2012modeling}
\bibinfo{author}{B.~E. Treeby}, \bibinfo{author}{J.~Jaros},
  \bibinfo{author}{A.~P. Rendell}, and \bibinfo{author}{B.~Cox},
  \enquote{\bibinfo{title}{Modeling nonlinear ultrasound propagation in
  heterogeneous media with power law absorption using a k-space pseudospectral
  method}},  \bibinfo{journal}{The Journal of the Acoustical Society of
  America} \textbf{131}(6), \bibinfo{pages}{4324--4336} (\bibinfo{year}{2012}).

\bibitem{cox2018accurate}
\bibinfo{author}{B.~Cox} and \bibinfo{author}{B.~Treeby},
  \enquote{\bibinfo{title}{Accurate time-varying sources in k-space
  pseudospectral time domain acoustic simulations}}, in
  \emph{\bibinfo{booktitle}{2018 IEEE International Ultrasonics Symposium
  (IUS)}}, \bibinfo{organization}{IEEE} (\bibinfo{year}{2018}), pp.
  \bibinfo{pages}{1--4}.

\bibitem{treeby2020nonlinear}
\bibinfo{author}{B.~E. Treeby}, \bibinfo{author}{E.~S. Wise},
  \bibinfo{author}{F.~Kuklis}, \bibinfo{author}{J.~Jaros}, and
  \bibinfo{author}{B.~Cox}, \enquote{\bibinfo{title}{Nonlinear ultrasound
  simulation in an axisymmetric coordinate system using a k-space
  pseudospectral method}},  \bibinfo{journal}{The Journal of the Acoustical
  Society of America} \textbf{148}(4), \bibinfo{pages}{2288--2300}
  (\bibinfo{year}{2020}).

\bibitem{jaros2016full}
\bibinfo{author}{J.~Jaros}, \bibinfo{author}{A.~P. Rendell}, and
  \bibinfo{author}{B.~E. Treeby}, \enquote{\bibinfo{title}{Full-wave nonlinear
  ultrasound simulation on distributed clusters with applications in
  high-intensity focused ultrasound}},  \bibinfo{journal}{The International
  Journal of High Performance Computing Applications} \textbf{30}(2),
  \bibinfo{pages}{137--155} (\bibinfo{year}{2016}).

\bibitem{gu2021msound}
\bibinfo{author}{J.~Gu} and \bibinfo{author}{Y.~Jing},
  \enquote{\bibinfo{title}{{mSOUND}: An open source toolbox for modeling
  acoustic wave propagation in heterogeneous media}},  \bibinfo{journal}{IEEE
  Transactions on Ultrasonics, Ferroelectrics, and Frequency Control}
  \textbf{68}(5), \bibinfo{pages}{1476--1486} (\bibinfo{year}{2021}).

\bibitem{wout2015fast}
\bibinfo{author}{E.~van~'t Wout}, \bibinfo{author}{P.~G{\'e}lat},
  \bibinfo{author}{T.~Betcke}, and \bibinfo{author}{S.~Arridge},
  \enquote{\bibinfo{title}{A fast boundary element method for the scattering
  analysis of high-intensity focused ultrasound}},  \bibinfo{journal}{The
  Journal of the Acoustical Society of America} \textbf{138}(5),
  \bibinfo{pages}{2726--2737} (\bibinfo{year}{2015}).

\bibitem{wout2022highcontrast}
\bibinfo{author}{E.~van~'t Wout}, \bibinfo{author}{S.~R. Haqshenas},
  \bibinfo{author}{P.~G{\'e}lat}, \bibinfo{author}{T.~Betcke}, and
  \bibinfo{author}{N.~Saffari}, \enquote{\bibinfo{title}{Boundary integral
  formulations for acoustic modelling of high-contrast media}},
  \bibinfo{journal}{Computers \& Mathematics with Applications} \textbf{105},
  \bibinfo{pages}{136--149} (\bibinfo{year}{2022}).

\bibitem{haqshenas2021fast}
\bibinfo{author}{S.~R. Haqshenas}, \bibinfo{author}{P.~G{\'e}lat},
  \bibinfo{author}{E.~van~'t Wout}, \bibinfo{author}{T.~Betcke}, and
  \bibinfo{author}{N.~Saffari}, \enquote{\bibinfo{title}{A fast full-wave
  solver for calculating ultrasound propagation in the body}},
  \bibinfo{journal}{Ultrasonics} \textbf{110}, \bibinfo{pages}{106240}
  (\bibinfo{year}{2021}).

\bibitem{wout2021benchmarking}
\bibinfo{author}{E.~van~'t Wout}, \bibinfo{author}{S.~R. Haqshenas},
  \bibinfo{author}{P.~G{\'e}lat}, \bibinfo{author}{T.~Betcke}, and
  \bibinfo{author}{N.~Saffari}, \enquote{\bibinfo{title}{Benchmarking
  preconditioned boundary integral formulations for acoustics}},
  \bibinfo{journal}{International Journal for Numerical Methods in Engineering}
  \textbf{122}(20), \bibinfo{pages}{5873--5897} (\bibinfo{year}{2021}).

\bibitem{betcke2017computationally}
\bibinfo{author}{T.~Betcke}, \bibinfo{author}{E.~van~'t Wout}, and
  \bibinfo{author}{P.~G{\'e}lat}, \enquote{\bibinfo{title}{Computationally
  efficient boundary element methods for high-frequency {H}elmholtz problems in
  unbounded domains}}, in \emph{\bibinfo{booktitle}{Modern Solvers for
  {H}elmholtz Problems}},  edited by \bibinfo{editor}{D.~Lahaye},
  \bibinfo{editor}{J.~Tang}, and \bibinfo{editor}{K.~Vuik}, Geosystems
  Mathematics  (\bibinfo{publisher}{Birkh\"auser}, \bibinfo{address}{Cham},
  \bibinfo{year}{2017}), pp. \bibinfo{pages}{215--243}.

\bibitem{wout2021pmchwt}
\bibinfo{author}{E.~van~'t Wout}, \bibinfo{author}{S.~Haqshenas},
  \bibinfo{author}{P.~G{\'e}lat}, \bibinfo{author}{T.~Betcke}, and
  \bibinfo{author}{N.~Saffari}, \enquote{\bibinfo{title}{Frequency-robust
  preconditioning of boundary integral equations for acoustic transmission}},
  (\bibinfo{year}{2021}), \bibinfo{note}{arXiv Preprint Arxiv:2104.04609}.

\bibitem{smigaj2015solving}
\bibinfo{author}{W.~{\'S}migaj}, \bibinfo{author}{T.~Betcke},
  \bibinfo{author}{S.~Arridge}, \bibinfo{author}{J.~Phillips}, and
  \bibinfo{author}{M.~Schweiger}, \enquote{\bibinfo{title}{Solving boundary
  integral problems with {BEM++}}},  \bibinfo{journal}{ACM Transactions on
  Mathematical Software (TOMS)} \textbf{41}(2), \bibinfo{pages}{6}
  (\bibinfo{year}{2015}).

\bibitem{geuzaine2009gmsh}
\bibinfo{author}{C.~Geuzaine} and \bibinfo{author}{J.-F. Remacle},
  \enquote{\bibinfo{title}{Gmsh: A {3-D} finite element mesh generator with
  built-in pre- and post-processing facilities}},
  \bibinfo{journal}{International Journal for Numerical Methods in Engineering}
  \textbf{79}(11), \bibinfo{pages}{1309--1331} (\bibinfo{year}{2009}).

\bibitem{meshmixer}
\bibinfo{author}{R.~Schmidt} and \bibinfo{author}{K.~Singh},
  \enquote{\bibinfo{title}{Meshmixer: An interface for rapid mesh
  composition}}, in \emph{\bibinfo{booktitle}{ACM SIGGRAPH 2010 Talks}},
  SIGGRAPH '10, \bibinfo{publisher}{Association for Computing Machinery},
  \bibinfo{address}{New York, NY, USA} (\bibinfo{year}{2010}).

\bibitem{afanasiev2019modular}
\bibinfo{author}{M.~Afanasiev}, \bibinfo{author}{C.~Boehm},
  \bibinfo{author}{M.~van Driel}, \bibinfo{author}{L.~Krischer},
  \bibinfo{author}{M.~Rietmann}, \bibinfo{author}{D.~A. May},
  \bibinfo{author}{M.~G. Knepley}, and \bibinfo{author}{A.~Fichtner},
  \enquote{\bibinfo{title}{Modular and flexible spectral-element waveform
  modelling in two and three dimensions}},  \bibinfo{journal}{Geophysical
  Journal International} \textbf{216}(3), \bibinfo{pages}{1675--1692}
  (\bibinfo{year}{2019}).

\bibitem{ferroni2017dispersion}
\bibinfo{author}{A.~Ferroni}, \bibinfo{author}{P.~F. Antonietti},
  \bibinfo{author}{I.~Mazzieri}, and \bibinfo{author}{A.~Quarteroni},
  \enquote{\bibinfo{title}{Dispersion-dissipation analysis of {3-D} continuous
  and discontinuous spectral element methods for the elastodynamics equation}},
   \bibinfo{journal}{Geophysical Journal International} \textbf{211}(3),
  \bibinfo{pages}{1554--1574} (\bibinfo{year}{2017}).

\bibitem{hapla2021fully}
\bibinfo{author}{V.~Hapla}, \bibinfo{author}{M.~G. Knepley},
  \bibinfo{author}{M.~Afanasiev}, \bibinfo{author}{C.~Boehm},
  \bibinfo{author}{M.~van Driel}, \bibinfo{author}{L.~Krischer}, and
  \bibinfo{author}{A.~Fichtner}, \enquote{\bibinfo{title}{Fully parallel mesh
  {I/O} using {PETSc DMPlex} with an application to waveform modeling}},
  \bibinfo{journal}{SIAM Journal on Scientific Computing} \textbf{43}(2),
  \bibinfo{pages}{C127--C153} (\bibinfo{year}{2021}).

\bibitem{kosloff1986absorbing}
\bibinfo{author}{R.~Kosloff} and \bibinfo{author}{D.~Kosloff},
  \enquote{\bibinfo{title}{Absorbing boundaries for wave propagation
  problems}},  \bibinfo{journal}{Journal of Computational Physics}
  \textbf{63}(2), \bibinfo{pages}{363--376} (\bibinfo{year}{1986}).

\bibitem{Note1}
\bibinfo{note}{Coreform Cubit (Version 2021.5) [Computer software]. Orem, UT:
  Coreform LLC. Retrieved from http://coreform.com}.

\bibitem{witte2019compressive}
\bibinfo{author}{P.~A. Witte}, \bibinfo{author}{M.~Louboutin},
  \bibinfo{author}{F.~Luporini}, \bibinfo{author}{G.~J. Gorman}, and
  \bibinfo{author}{F.~J. Herrmann}, \enquote{\bibinfo{title}{Compressive
  least-squares migration with on-the-fly {Fourier} transforms}},
  \bibinfo{journal}{Geophysics} \textbf{84}(5), \bibinfo{pages}{R655--R672}
  (\bibinfo{year}{2019}).

\bibitem{chew19943d}
\bibinfo{author}{W.~C. Chew} and \bibinfo{author}{W.~H. Weedon},
  \enquote{\bibinfo{title}{A {3D} perfectly matched medium from modified
  maxwell's equations with stretched coordinates}},
  \bibinfo{journal}{Microwave and Optical Technology Letters} \textbf{7}(13),
  \bibinfo{pages}{599--604} (\bibinfo{year}{1994}).

\bibitem{kyriakou2015full}
\bibinfo{author}{A.~Kyriakou}, \bibinfo{author}{E.~Neufeld},
  \bibinfo{author}{B.~Werner}, \bibinfo{author}{G.~Sz{\'e}kely}, and
  \bibinfo{author}{N.~Kuster}, \enquote{\bibinfo{title}{Full-wave acoustic and
  thermal modeling of transcranial ultrasound propagation and investigation of
  skull-induced aberration correction techniques: a feasibility study}},
  \bibinfo{journal}{Journal of Therapeutic Ultrasound} \textbf{3}(1),
  \bibinfo{pages}{1--18} (\bibinfo{year}{2015}).

\bibitem{neufeld2016approach}
\bibinfo{author}{E.~Neufeld}, \bibinfo{author}{A.~Kyriacou},
  \bibinfo{author}{W.~Kainz}, and \bibinfo{author}{N.~Kuster},
  \enquote{\bibinfo{title}{Approach to validate simulation-based distribution
  predictions combining the gamma-method and uncertainty assessment:
  application to focused ultrasound}},  \bibinfo{journal}{Journal of
  Verification, Validation and Uncertainty Quantification} \textbf{1}(3)
  (\bibinfo{year}{2016}).

\bibitem{pasquinelli2020transducer}
\bibinfo{author}{C.~Pasquinelli}, \bibinfo{author}{H.~Montanaro},
  \bibinfo{author}{H.~J. Lee}, \bibinfo{author}{L.~G. Hanson},
  \bibinfo{author}{H.~Kim}, \bibinfo{author}{N.~Kuster}, \bibinfo{author}{H.~R.
  Siebner}, \bibinfo{author}{E.~Neufeld}, and \bibinfo{author}{A.~Thielscher},
  \enquote{\bibinfo{title}{Transducer modeling for accurate acoustic
  simulations of transcranial focused ultrasound stimulation}},
  \bibinfo{journal}{Journal of Neural Engineering} \textbf{17}(4),
  \bibinfo{pages}{046010} (\bibinfo{year}{2020}).

\bibitem{montanaro2021impact}
\bibinfo{author}{H.~Montanaro}, \bibinfo{author}{C.~Pasquinelli},
  \bibinfo{author}{H.~J. Lee}, \bibinfo{author}{H.~Kim}, \bibinfo{author}{H.~R.
  Siebner}, \bibinfo{author}{N.~Kuster}, \bibinfo{author}{A.~Thielscher}, and
  \bibinfo{author}{E.~Neufeld}, \enquote{\bibinfo{title}{The impact of ct image
  parameters and skull heterogeneity modeling on the accuracy of transcranial
  focused ultrasound simulations}},  \bibinfo{journal}{Journal of Neural
  Engineering} \textbf{18}(4), \bibinfo{pages}{046041} (\bibinfo{year}{2021}).

\bibitem{cueto2021stride}
\bibinfo{author}{C.~Cueto}, \bibinfo{author}{O.~Bates},
  \bibinfo{author}{G.~Strong}, \bibinfo{author}{J.~Cudeiro},
  \bibinfo{author}{F.~Luporini}, \bibinfo{author}{O.~C. Agudo},
  \bibinfo{author}{G.~Gorman}, \bibinfo{author}{L.~Guasch}, and
  \bibinfo{author}{M.-X. Tang}, \enquote{\bibinfo{title}{Stride: a flexible
  platform for high-performance ultrasound computed tomography}},
  \bibinfo{journal}{arXiv preprint arXiv:2110.03345}  (\bibinfo{year}{2021}).

\bibitem{amundsen2017time}
\bibinfo{author}{L.~Amundsen} and \bibinfo{author}{{\O}.~Pedersen},
  \enquote{\bibinfo{title}{Time step n-tupling for wave equations}},
  \bibinfo{journal}{Geophysics} \textbf{82}(6), \bibinfo{pages}{T249--T254}
  (\bibinfo{year}{2017}).

\bibitem{yao2018effective}
\bibinfo{author}{G.~Yao}, \bibinfo{author}{N.~V. Da~Silva}, and
  \bibinfo{author}{D.~Wu}, \enquote{\bibinfo{title}{An effective absorbing
  layer for the boundary condition in acoustic seismic wave simulation}},
  \bibinfo{journal}{Journal of Geophysics and Engineering} \textbf{15}(2),
  \bibinfo{pages}{495--511} (\bibinfo{year}{2018}).

\bibitem{gao2015unsplit}
\bibinfo{author}{Y.~Gao}, \bibinfo{author}{J.~Zhang}, and
  \bibinfo{author}{Z.~Yao}, \enquote{\bibinfo{title}{Unsplit complex frequency
  shifted perfectly matched layer for second-order wave equation using
  auxiliary differential equations}},  \bibinfo{journal}{The Journal of the
  Acoustical Society of America} \textbf{138}(6), \bibinfo{pages}{EL551--EL557}
  (\bibinfo{year}{2015}).

\bibitem{hicks2002arbitrary}
\bibinfo{author}{G.~J. Hicks}, \enquote{\bibinfo{title}{Arbitrary source and
  receiver positioning in finite-difference schemes using {Kaiser} windowed
  sinc functions}},  \bibinfo{journal}{Geophysics} \textbf{67}(1),
  \bibinfo{pages}{156--165} (\bibinfo{year}{2002}).

\bibitem{martin2019investigation}
\bibinfo{author}{E.~Martin} and \bibinfo{author}{B.~Treeby},
  \enquote{\bibinfo{title}{Investigation of the repeatability and
  reproducibility of hydrophone measurements of medical ultrasound fields}},
  \bibinfo{journal}{The Journal of the Acoustical Society of America}
  \textbf{145}(3), \bibinfo{pages}{1270--1282} (\bibinfo{year}{2019}).

\end{thebibliography}



\end{document}